\documentclass[10pt,journal,compsoc, twoside]{IEEEtran}
\ifCLASSOPTIONcompsoc
  \usepackage[nocompress]{cite}
\else
  \usepackage{cite}
\fi
\ifCLASSINFOpdf
\else
\fi
 \usepackage{listings}
 \usepackage{graphicx}
 \usepackage{subfig}
 \usepackage{xcolor}
 \usepackage{textcomp}
 \usepackage{amsmath}
 \usepackage{amsfonts}
 \usepackage{url}
 \usepackage{multirow}
 \usepackage{prettyref}
 \newrefformat{eqn}{Equation~\ref{#1}}
 \newrefformat{fig}{Figure~\ref{#1}}
 \newrefformat{tab}{Table~\ref{#1}}
 \newrefformat{sec}{Section~\ref{#1}}
 \newrefformat{lst}{Listing~\ref{#1}}

\begin{document}
%\title{An Adaptive Sampling Method for Monte Carlo Simulation Targeting an HPC-based Car Navigation System}
%\title{An Efficient Probabilistic Time Dependent Routing Calculation Targeting an HPC-based Car Navigation System}
\title{An Efficient Monte Carlo-based Probabilistic Time-Dependent Routing Calculation Targeting a Server-Side Car Navigation System}

%\author{Michael~Shell,~\IEEEmembership{Member,~IEEE,}
        %John~Doe,~\IEEEmembership{Fellow,~OSA,}
        %and~Jane~Doe,~\IEEEmembership{Life~Fellow,~IEEE}% <-this % stops a space
        
\author{Emanuele Vitali, Davide Gadioli, Gianluca Palermo, \\ 
Martin Golasowski, Jo\~{a}o Bispo, Pedro Pinto, Jan Martinovi\v{c}, Kate\v{r}ina Slaninov\'{a}, \\
Jo\~{a}o M. P. Cardoso, Cristina Silvano

\IEEEcompsocitemizethanks{
\IEEEcompsocthanksitem E. Vitali, D. Gadioli, G. Palermo, and C. Silvano are with Dipartimento di Elettronica, Infomazione e Bioingegneria, Politecnico di Milano, Italy.
\IEEEcompsocthanksitem J. Bispo, P. Pinto and J. Cardoso are with Faculty of Engineering (FEUP), University of Porto, Portugal.
\IEEEcompsocthanksitem Martin Golasowski, Jan Martinovi\v{c} and Kate\v{r}ina Slaninov\'{a}, are with IT4Innovations, V\v{S}B - Technical University of Ostrava, Czech Republic.
%E-mail: see http://www.michaelshell.org/contact.html
%\IEEEcompsocthanksitem J. Doe and J. Doe are with Anonymous University.\protect\\
}% <-this % stops an unwanted space
%\thanks{Manuscript received Month Day, 2018; revised August 26, 2015.}
%\thanks{This work is supported by the European Union's Horizon 2020 research and innovation program under grant agreement No 671623, FET-HPC ANTAREX.}
%\thanks{Manuscript received July 31, 2018; revised Month XX, 201X.}
}
        
%\IEEEcompsocitemizethanks{\IEEEcompsocthanksitem M. Shell was with the Department
%of Electrical and Computer Engineering, Georgia Institute of Technology, Atlanta,
%GA, 30332.\protect\\
%E-mail: see http://www.michaelshell.org/contact.html
%\IEEEcompsocthanksitem J. Doe and J. Doe are with Anonymous University.}% <-this % stops an unwanted space

\lstset{numbers=left,xleftmargin=2em, basicstyle=\scriptsize}

\IEEEtitleabstractindextext{%
\begin{abstract}

Incorporating speed probability distribution to the computation of the route planning in car navigation systems guarantees more accurate and precise responses.
In this paper, we propose a novel approach for dynamically selecting the number of samples used for the Monte Carlo simulation to solve the Probabilistic Time-Dependent Routing (PTDR) problem, thus improving the computation efficiency. 
The proposed method is used to determine in a proactive manner the number of simulations to be done to extract the travel-time estimation for each specific request while respecting an error threshold as output quality level.
The methodology requires a reduced effort on the application development side. We adopted an aspect-oriented programming language (LARA) together with a flexible dynamic autotuning library (mARGOt) respectively to instrument the code and to take tuning decisions on the number of samples improving the execution efficiency.
Experimental results demonstrate that the proposed adaptive approach saves a large fraction of simulations (between 36\% and 81\%) with respect to a static approach while considering different traffic situations, paths and error requirements. Given the negligible runtime overhead of the proposed approach, it results in an execution-time speedup between 1.5x and 5.1x. 
This speedup is reflected at infrastructure-level in terms of a reduction of around 36\% of the computing resources needed to support the whole navigation pipeline.

\end{abstract}

}

\maketitle

\IEEEdisplaynontitleabstractindextext

% \fi
\IEEEpeerreviewmaketitle

\section{Introduction}

\IEEEPARstart{I}{n} smart cities, the trend is to combine and automate several common tasks to ease the life of citizens. 
Among these tasks, traffic estimation and prediction plays a central role: it is used not only to avoid traffic congestion, which allows to have predictable travel times, but also to reduce car emissions. 
Considering the rising wave of self-driving cars, the amount of car navigation requests will increase rapidly together with the need of real-time updates and processing on large graphs representing the urban network. 
This trend imposes larger and more powerful computing infrastructures composed of HPC resources.

Concerning the algorithmic problem, car navigation is one of the main problems of applied theoretical research. The Dijsktra's shortest path algorithm is used for finding the optimal path between two vertices in a weighted graph representing a road network. Apart from single navigation between two points, navigation algorithms are used in various systems for solving larger optimization problems, like route planning for a fleet of package delivery vehicles, waste collection management or traffic optimization in a smart city \cite{toth2014vehicle}. 
Definition of the optimal path is based on the type of used weights of the graph edges. The shortest path is based on the geographical distance between two adjacent vertices of a graph. The fastest path is based on time needed to cross a particular edge. There might be more complex criteria, however their description is out of the scope of this paper. Time needed to cross a particular stretch of road can be affected by various elements, such as accidents, traffic congestion, road work and so on. At the basic level, upper legal limit of speed is used, based on the assumption that each vehicle travels with the same speed. This can be vastly inaccurate due to the natural behavior of traffic. 

With increasing availability of historical traffic monitoring data, there are several research efforts to determine average speed on road networks by using statistical analysis and various models. However, a single speed value is still not very useful as it does not reflect the stochastic behavior of the traffic. The probability distribution of the speed at a certain time enables to incorporate low probability real world events that can cause major delays and affect traffic over vast areas. By incorporating probability distribution to the computation, the system can compute the probability of arrival time within a certain time-frame which can be useful for more precise route planning. This problem is called \emph{Probabilistic Time-Dependent Routing} (PTDR).

A scalable algorithm for solving the PTDR problem based on Monte Carlo simulations has been presented in \cite{tomis2015probabilistic}\cite{golasowski2016performance} and represents the base for our work. 
In particular, the algorithm uses probability distributions of travel time for the individual graph edges to estimate the distribution of the total travel time and it is integrated to an experimental server-side routing service. This service is deployed on an HPC infrastructure to offer optimal performance for a large number of requests as needed by the smart city context. 
The PTDR algorithm employed in this work simulates a large number of vehicles driving along a determined path in a graph at a particular time of departure. The speed of vehicles on individual roads is sampled from the speed probability distribution (also called speed profile) associated to the graph edge. The number of samples is a parameter that directly affects the informational value of the output as well as its computational requirements. Given the large amount of requests to be served, 
even small changes in the workload can affect the overall HPC system efficiency. 
While the original version was based on a worst-case tuning of the number of samples \cite{tomis2015probabilistic}, and given that a reactive approach \cite{Gilman:1968:BSS:800166.805256} is not a viable solution due to the overheads, in this paper we present a proactive method for dynamically adapting the number of samples for the Monte Carlo (MC) based PTDR algorithm.

In particular, the main contributions of this paper can be summarized as follows:
\begin{itemize}
\item We propose a methodology for self-adapting the PTDR algorithm presented in \cite{tomis2015probabilistic}\cite{golasowski2016performance} to the input data in a proactive manner, maximizing its performance while respecting the output quality level;  
\item We propose a probabilistic error model used to correlate the input data characteristics with the number of samples used by the Monte Carlo algorithm; 
\item We adopted an aspect-oriented programming language to keep separated the functional version of the application from the code needed to introduce the adaptivity layer. 
\end{itemize}

The remainder of this paper is organized as follows.
\prettyref{sec:sota} provides an overview of the related works, while \prettyref{sec:MC} provides an introduction to the Monte Carlo approach to solve the PTDR problem. \prettyref{sec:proposedsolution} and \prettyref{sec:integ} describe the proposed methodology respectively from the adaptivity and code integration point of view. 
Finally, \prettyref{sec:experiment} describes the experimental campaign to validate the proposed methodology and \prettyref{sec:conc} concludes the paper.

\section{Related Work}
\label{sec:sota}
Determining the optimal path in a stochastic time-dependent graph is a well-studied problem which has many formulations \cite{agafonov2016reliable}. Our approach is closest to the \textit{Shortest-path problem with on-time
arrival reliability} (SPOTAR) formulation. It can be seen as a variant of the \textit{Stochastic on-time arrival} (SOTA) problem, for which practical solution exists as shown in \cite{samaranayake2011tractable}. These algorithms have the objective of maximizing the probability of arriving within a time budget and are related to optimal routing in stochastic networks. However, there are not many solutions for the time-dependent variant of both of the problems. In \cite{agafonov2016reliable} authors show practical results for the time-dependent variant of SOTA, simultaneously in \cite{nie2009shortest} authors elaborate on the complexity of existing theoretical solutions of the SPOTAR problem and show how it can be extended with time dependency. There are many other papers which show various theoretical approaches for the SOTA problem, including some practical applications \cite{samaranayake2011tractable}\cite{abeydeera2014gpu}\cite{niknami2016tractable}\cite{nikolova2006stochastic}. Solution to the SPOTAR problem based on \emph{policy-based} SOTA as a heuristic is presented \cite{niknami2016tractable}. However, the authors make the assumption that the network is time-invariant, which is not true in real cases if considering long paths. The solution is also unusable in on-line systems as its scalability to graphs representing real-world routes is not sufficient.

Our approach follows the same philosophy presented in \cite{tomis2015probabilistic, golasowski2016performance} where the authors provides an approximate solution of the time-dependent variant of the SPOTAR problem based on Monte Carlo simulations. As shown in Section \ref{sec:MC}, our approach uses the k-shortest paths algorithm \cite{IARP}\cite{Theodoros15Alternative}\cite{Theodoros17Exact} to determine the paths for which the travel time distribution is estimated. This separation allows us to implement the approach in an online system which provides adaptive routing in real-time.
Given the Monte Carlo nature of the algorithm, to improve the efficiency of the PTDR calculation we have two main alternatives \cite{JANSSEN2013123}. The first is the sampling efficiency, while the second is the sampling convergence.
In both cases, the algorithm optimization is reached by exploiting the iterative nature of the Monte Carlo simulation.
To obtain the sampling efficiency, several techniques have been proposed to determine what is the next sample to be evaluated to maximize the gathered knowledge \cite{4407879}, \cite{JANSSEN2013123}. However, in the implementation under analysis this has been discarded because our goal is to exploit the parallelism of the underlying HPC architecture \cite{golasowski2016performance} that excludes any iterative approach to the Monte Carlo.
For the same reason also the approaches that require a statistical property evaluation after every iteration \cite{Gilman:1968:BSS:800166.805256}, checking if the error is acceptable, cannot be considered acceptable.
Both approaches would be too time-consuming and, how it has been already analyzed in \cite{tomis2015probabilistic}, for the specific problem the number of samples has to be chosen a priori in a proactive rather than in a reactive manner.

In this paper, we adopted autotuning techniques to face with the Monte Carlo efficiency problem. 
Classical autotuning approaches are based on code refactoring and loop parameterization to optimize extra-functional properties (e.g. power and performance) on a given target architecture
(e.g. A-tune\cite{10.1007/978-3-642-03869-3_5}, SPIRAL\cite{Moura:00}, ATLAS\cite{Whaley:1998:ATL:509058.509096}, PowerDial \cite{Hoffmann:2011:DKR:1950365.1950390}).
More related to a Monte Carlo application, in \cite{6449942} the authors developed a framework capable to automatically select for a Monte Carlo application the optimal parameters for GPU accelerated kernels, e.g. block size, thread numbers and other variables of the CUDA runtime. 
All those approaches are orthogonal to the proposed work that is more focused on reducing the number of samples, rather than tailoring the computation for a specific platform.

Two interesting works facing the autotuning problem considering the output accuracy, also on Monte Carlo simulations, are those presented in \cite{Vassiliadis2016} and \cite{5764677}. In both cases, they employ an ad-hoc DSL language as front-end for the developer, letting the compilation framework to customize the code and related execution.
In particular, in \cite{Vassiliadis2016}, the authors propose a framework for driving the choice of extra-functional properties like the number of required processors, frequency and computation accuracy targeting the energy minimization. More related to tailoring the computation to the specific platform is the work in \cite{5764677} where the authors enhanced the framework to include compile time performance-accuracy trade-off.
The proposed approach tries to face the problem from a different perspective. First, we intend to reduce as much as possible the intrusiveness on the target code, thus reducing also the load on the application developer, and second, our approach relies more on run-time and data-aware decisions, also avoiding compile-time decisions on performance-accuracy trade-offs. 

A two-step approach for solving the Monte Carlo problem has been envisioned in \cite{7551421}. Similar to our work, the authors suggest to have a first shot of a reduced number of samples to provide an initial approximate solution as fast as possible, and then to refine the output to the required accuracy in successive iterations. In the proposed context, this idea suffers from two main problems. First, it is suitable for scientific work-flows where an intermediate solution is used to trigger next computations, and it is not our case. Second, in the iterative phase, it suggests a reactive approach rather than a proactive one, that we already discussed to be necessary for the specific PTDR problem.
Another two step approach is the one presented in \cite{Laurenzano:2016:IRU:2908080.2908087}.
In this case, the authors suggest building from each input a \emph{small canary} -- a statistically representative subset of the actual input -- that is used to perform a parameter exploration at runtime.
In this specific PTDR case, the approach suffers from a large overhead due to the canary extraction and approximation selection. In the proposed approach, we moved the two operations at design time by generalizing the approximation selection according to the \emph{unpredictability function} computed at runtime. Moreover, the proposed approach has been designed to statistically guarantee the respect of an output error level. 

Finally in \cite{Sui:2016:PCA:2872362.2872402}, the authors present a proactive control of the application tuning as a constrained optimization problem that can be solved at run-time according to profiling information. The authors suggest using a Bayesian model to learn the accuracy of the computation according to data features that are mainly related to the data size. Despite the similarity of this previous work with one of the components we used for the integration (i.e. the mARGOt autotuner), the proposed approach is mainly related to the data-aware error prediction model for PTDR, where SLA constraints are considered, and to the seamless integration of the adaptivity concepts.

\section{Monte Carlo Approach for Probabilistic Time-Dependent Routing}
\label{sec:MC}

So far, many theoretical formulations and several algorithms have been developed for solving the problem of computing the travel time distribution  \cite{agafonov2016reliable}. 
In this paper we consider a \emph{path-based} approach (\emph{SPOTAR}) where the paths are known \emph{a-priori} and travel-time distributions are determined subsequently for each one of the paths \cite{miller2003path}. 

\begin{figure}[t]
	\includegraphics[width=\columnwidth]{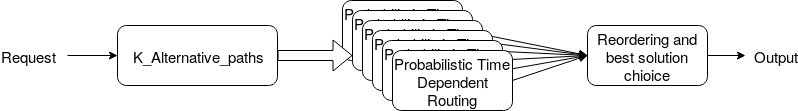}
    \caption{The complete navigation infrastructure for serving a single request.}
    \label{fig:infrastruct}
\end{figure}

In the context of the complete traffic navigator application illustrated in \prettyref{fig:infrastruct}, our focus is on the efficient estimation of the arrival time distribution (\emph{PTDR - Probabilistic Time-Dependent Routing} phase).
More in detail, the three main steps of the application can be described as follows:
(i) the first step consists of determining \emph{K alternative paths} to be passed to the next steps. In the navigation scenario, the identification of the shortest path is not enough to determine a good solution, if no traffic information have been considered. Thus alternative routes derived by algorithms for determining \emph{k-Short Paths} with limited overlap have to be adopted in this step \cite{Theodoros15Alternative}\cite{Theodoros17Exact}\cite{IARP}. This first phase is out of the scope of this paper;
(ii) For every path selected by the previous step (K-alternative paths), the computation of the travel time is done using the Probabilistic Time-Dependent Routing module. While the exact solution to the travel-time estimation (PTDR) has an exponential complexity, in this work we efficiently approximate the solution of the SPOTAR problem by adopting a Monte Carlo sampling approach \cite{tomis2015probabilistic};
(iii) The final step gathers the timing information provided by the $k$ instances of the PTDR module for every single request and selects the best path to be given back to the user. Actually, this phase does not provide a single route but reorders the list of $k$ paths determined by the first step according to the timing distributions determined in the second phase and user preference \cite{jan10sort}. 

This three-step approach of the whole navigation application allows us to implement an approximate solution to the SPOTAR problem, which can be used online in a system to serve a large volume of routing requests.

\label{sec:MCSapproach}

Our definition of a probabilistic road network is similar to the definition of the stochastic time-dependent network as described by Miller-Hooks \cite{miller2003path}, with the exception of the segment travel times, which has been substituted by the speed probability distribution (\emph{speed profile}) for a given time of departure within a week. Formally, it can be defined as follows. 
Let $ G = (V,E) $ be a well connected, directed and weighted graph, where $ V $ is the set of vertices and $ E $ is the set of edges. Each vertex represents a junction or some important point corresponding to geospatial properties of the road, while edges represent the individual road segments between the junctions. Each path selected by the first phase of the application (i.e. K-Alternative paths) can be formally represented as a vector of graph edges $ S = ( s_1, s_2, \dots, s_n) $, while $ S_p \subseteq E $ and $ n $ is the number of road segments in the path. 

Using a travel time estimation function, we are interested in estimating the travel time $\theta$ as $\hat{\theta}_{S,t,P_S}$ where $S$ is the given path, $ t $ is the departure times and $ P_S $ are the probabilistic speed profiles for the segments in $S$.
More in detail, $ t 
\in T $ is a departure time within a set of possible departure times which divide a certain timeframe to a set of intervals $ T = \{t: t = n \cdot \phi, n \in \mathbb{N}\} $
\cite{asghari2015probabilistic}, where the length of the interval $ \phi $ is determined by input data.
$ P $ is the set of probabilistic speed profiles for the entire graph edges $E$, where $ P_S \subseteq P $.
Each speed profile $ p \in P $ is represented by a set of discrete speed values and assigned probabilities. The number of speed values depends on the method used for deriving the profiles from historical traffic monitoring data, while the minimum and maximum values represent respectively the congestion speed and the free flow speed. 

In our work, the time frame is set to \textit{one week} and $ \phi = 900 s$ (15 minutes). This approach reflects traffic variations during the various hours of the day and for all the days of a week. By extending the time frame, other factors can be included, such as seasons or holidays. The number of speed values has been set to 4 levels according to the characteristics of the input data used for the creation of the speed profiles.

Focusing on the SPOTAR problem, we are not interested in a single travel time value $\theta$ but we require to calculate the probability distribution of the arrival time.
Given the previous formalization of the problem, 
the travel time distribution can be estimated by traversing the path segments together while considering the speed profile distribution.
In particular, we can define a tree where each layer represents a segment in the selected path \cite{tomis2015probabilistic}. The tree root is the starting segment, while the end segment is on the leaves. Each node in all layers of the tree has $ l $ children, where $ l $ is a number of the discrete speed values for each segment, and the tree depth corresponds to the number of selected path segments $ |S| $. Each edge in the tree is annotated by the discrete speed value, its probability, and by the length of the considered segment.
A travel time can be computed by a depth-first search (DFS) while selecting an arbitrary child node at each level of the tree. The travel time value is then the sum of the time spent in each segment (length/speed), while the probability of that value is the product of the probability on each edge of the traversal. 
Each traversal corresponds to a single car traveling along the entire path. 
The exact solution is obtained by an exhaustive search over all the possible paths between the root node and all the leaves.
This approach is clearly not efficient since it scales exponentially with the number of segments in the path.

A Monte Carlo-based approach can be successfully employed in this case. By generating a large number of random tree traversals, enough samples can be obtained to estimate the final distribution. We define this final distribution, which is a collection of $\theta$ values ($\theta_1 ... \theta_x$) obtained through the Monte Carlo simulation $MCS(x,i)$, where $x$ is the number of random tree traversals, and $i$ is the input set of the $\hat{\theta}$ function (i.e. $S, t, P_S$). 

Given that travel times usually have a long tailed distribution due to inherent properties of the traffic (e.g. rare events such as accidents) a large number of samples is needed to estimate the travel time distribution with a sufficient precision. 
Regarding the definition of the number of samples for the Monte Carlo simulation, the particular implementation of the PTDR kernel cannot rely on a run-time stability analysis of the output. Each tree traversal (a sample of the Monte Carlo simulation) is totally independent to the others, thus this problem is perfectly suitable for parallel computing architectures, such as modern CPUs or accelerators. To efficiently exploit this parallelism, it is necessary to know a-priori the number of travel time estimations required to build the final distribution.

\begin{figure}[t]
	\includegraphics[width=\columnwidth]{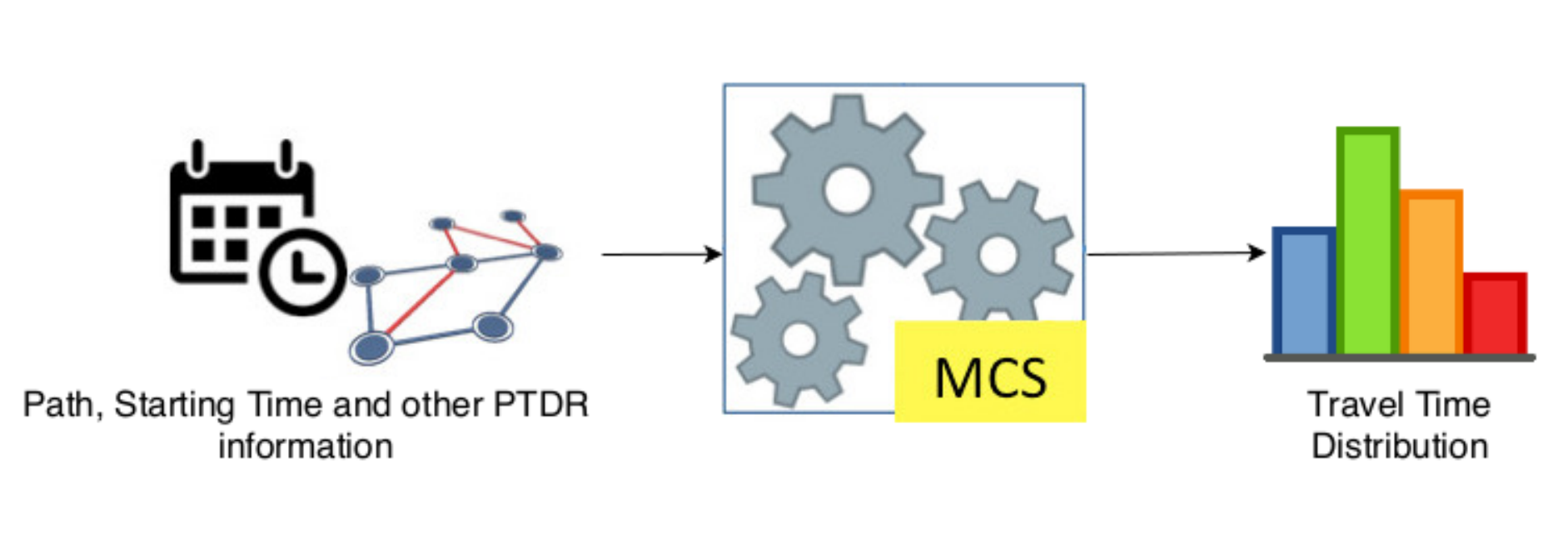}
    \caption{The original approach for PTDR routing based on Monte Carlo simulations to derive the travel time distribution.}
    \label{fig:original}
\end{figure}

To summarize, the PTDR algorithm can be seen as in \prettyref{fig:original}, where all the information regarding the request are provided to the Monte Carlo simulation (MCS) capable to return the predicted travel time distribution for the given route.

\section{The Proposed Approach}
\label{sec:proposedsolution}
The Monte Carlo simulation is designed to use a given number of samples $x$ for every run. Based on a conventional approach, this number is  selected according to the worst-case analysis and it is the lowest number of samples always able to reach a target precision \cite{golasowski2016performance}.
In this section, we present the proposed technique adopted to select at runtime the number of samples for the Monte Carlo simulation according to the input data characteristics. 

Before moving on the methodological part, let us better define the specific context of the problem. In particular, even if we are interested in the travel time distribution, our goal is to know a value $\tau_i$ to guarantee with a certain probability that the travel time will be within that value: $P(\theta<\tau_i)\geq y$ where $i$ has been defined as the input set of the travel-time function.
The value $\tau_i$ is the output of the PTDR phase. In the following, we characterize that value with an additional property $\tau_{i, y}$, where y is the probability that the travel time will be lower than $\tau$.

Using the Monte Carlo simulation, we can estimate the value of $\tau_{i, y}$ using $x$ samples as follows $\hat{\tau}_{i,y}^x = MCS(x,i,y)$. In particular, we estimate the value $\hat{\tau}_{i,y}^x$ by selecting the y-th percentile of the finite-sample distribution obtained from the Monte Carlo simulation (i.e. if $y=95\%$ then $\hat{\tau}_{i,y}^x$ is the $95^{th}$ percentile of the distribution).

In the context of this work, we are interested in minimizing the execution time of the function $MCS$, while limiting the prediction error defined as $error_{i,y}^x  = \frac{|\tau _{i,y} - \hat{\tau} _{i,y}^x|}{\tau _{i,y}}$.
In particular, the target problem can be expressed as follows: 
\begin{equation}
\begin{aligned}
& \underset{x}{\text{minimize}} & & cpu\_time_i^x \\
& \text{subject to} & & error_{i,y}^x \le \epsilon %, \; i = 1, \ldots, m.
%& & n \times \sigma_{x,i,y} \le \epsilon%, \; i = 1, \ldots, m.
\end{aligned}
\end{equation}
where $\epsilon$ represents the upper bound on the computation error.
We want this error to be relative to the output of the MCS, that is the desired percentile of the predicted travel time.
In this way, we can abstract from the actual path.
Given the tight correlation between the execution time and the used number of samples $x$, the previous problem can also be simplified by considering the minimization of $x$ instead of the $cpu\_time$. 
According to Monte Carlo properties, we can derive that $\tau _{i,y} \equiv \hat{\tau}_{i,y}^\infty$, where $\hat{\tau}_{i,y}^\infty$ is the output of the $MCS$ function computed using an infinite number of samples.
Thus, we can rewrite the error as 
\begin{equation}
error_{i,y}^x =\frac{|\hat{\tau} _{i,y}^\infty - \hat{\tau} _{i,y}^x|}{\hat{\tau} _{i,y}^\infty}
\end{equation}

Due to the Monte Carlo properties \cite{juritz1983accuracy}, the value $\hat{\tau} _{i,y}^x$ is a random variable, asymptotically normally distributed with mean $\mu_{\hat{\tau}_{i,y}^x}$ and standard deviation $\sigma_{\hat{\tau}_{i,y}^x}$.
In particular, according to the central limit theorem \cite{Montgomery2003AppliedStatistics}, while considering enough values of samples the mean value does not depend on the number of Monte Carlo simulations, and the standard deviation decreases by increasing the number of Monte Carlo simulations.
Given that, we can define the error as characterized by a normal distribution with mean $0$ and a standard deviation $\sigma_{\hat{\tau}_{i,y}^x}/\mu_{\hat{\tau}_{i,y}^x}$. 
In the following, we refer to the standard deviation of the error as $\nu_{\hat{\tau}_{i,y}^x} = \frac{\sigma_{\hat{\tau}_{i,y}^x}}{\mu_{\hat{\tau}_{i,y}^x}}$. This expression is the same as the coefficient of variation (relative standard deviation) of the result of the Monte Carlo simulation.

According to the probabilistic nature of the problem, we cannot guarantee that the error will be always below $\epsilon$. However, this can be done relaxing the error constraint by introducing a confidence interval (CI) level. In particular, given the normal distribution of the error, the selected confidence interval can be correlated with the expected error: 
\begin{equation}
P(error_{i,y}^x \leq \epsilon) \geq CI  \Longrightarrow
\hat{error}_{i,y}^x\leq n(CI) \times \nu_{\hat{\tau}_{i,y}^x} \leq \epsilon
\end{equation}
where $n(CI)$ is a value that express the confidence level (e.g. n(68\%)=1, n(95\%)=2 and n(99.7\%)=3 derived from  the 1-3 $\sigma$-intervals of the normal distribution). 
Thus, if we decrease the number of Monte Carlo simulations used to derive $\hat{\tau} _{i,y}^x$, on one hand we decrease the execution time of the application, but on the other hand we are also reducing the accuracy of the results, having a larger value for the coefficient of variation $\nu_{\hat{\tau}_{i,y}^x}$.

An additional problem is derived from the fact that $\hat{\tau}_{i,y}^x$ is input dependent. This means that it is not possible to predict the possible Monte Carlo error for unknown paths, according to the number of samples.
To deal with this, we found a feature $u_i$ of the inputs $i$ that can be used to quickly estimate the number of samples necessary to keep the error below the threshold $\epsilon$. The idea is to evaluate the error by using $u_i$ instead of the actual $i$ so that we can transform the original problem as 
\begin{equation}
error_{i,y}^x\leq n(CI) \times \nu_{\hat{\tau}_{u_i,y}^x}.
\end{equation}
The feature $u_i$ has been called \emph{unpredictability}, since it represents a set of characteristics of the inputs $i$ (road, starting time,...) that provides information about how complex is the prediction of $\tau_{i,y}$, therefore it is also related on how many samples are required to satisfy a certain error and confidence level.
More details on the unpredictability feature are presented in Section \ref{sec:unpredictability}.

Given that the error is not anymore related to the specific input set $i$ but only to the feature $u_i$, the number of samples needed to satisfy the constraint can be easily extracted by $\nu_{\hat{\tau}_{u_i,y}^x} \leq \frac{\epsilon}{n(CI)}$. A profiling phase on a set of representative inputs can be used to then extract the values of $\hat{\nu}_{\hat{\tau}_{u_i,y}^x}$, that will be used to determine the correlation between the unpredictability function and the error. 
More details on the profiling phase including the prediction function are presented in Section \ref{sec:unpredictability_profiling}.

\begin{figure}[t]
	\includegraphics[width=\columnwidth]{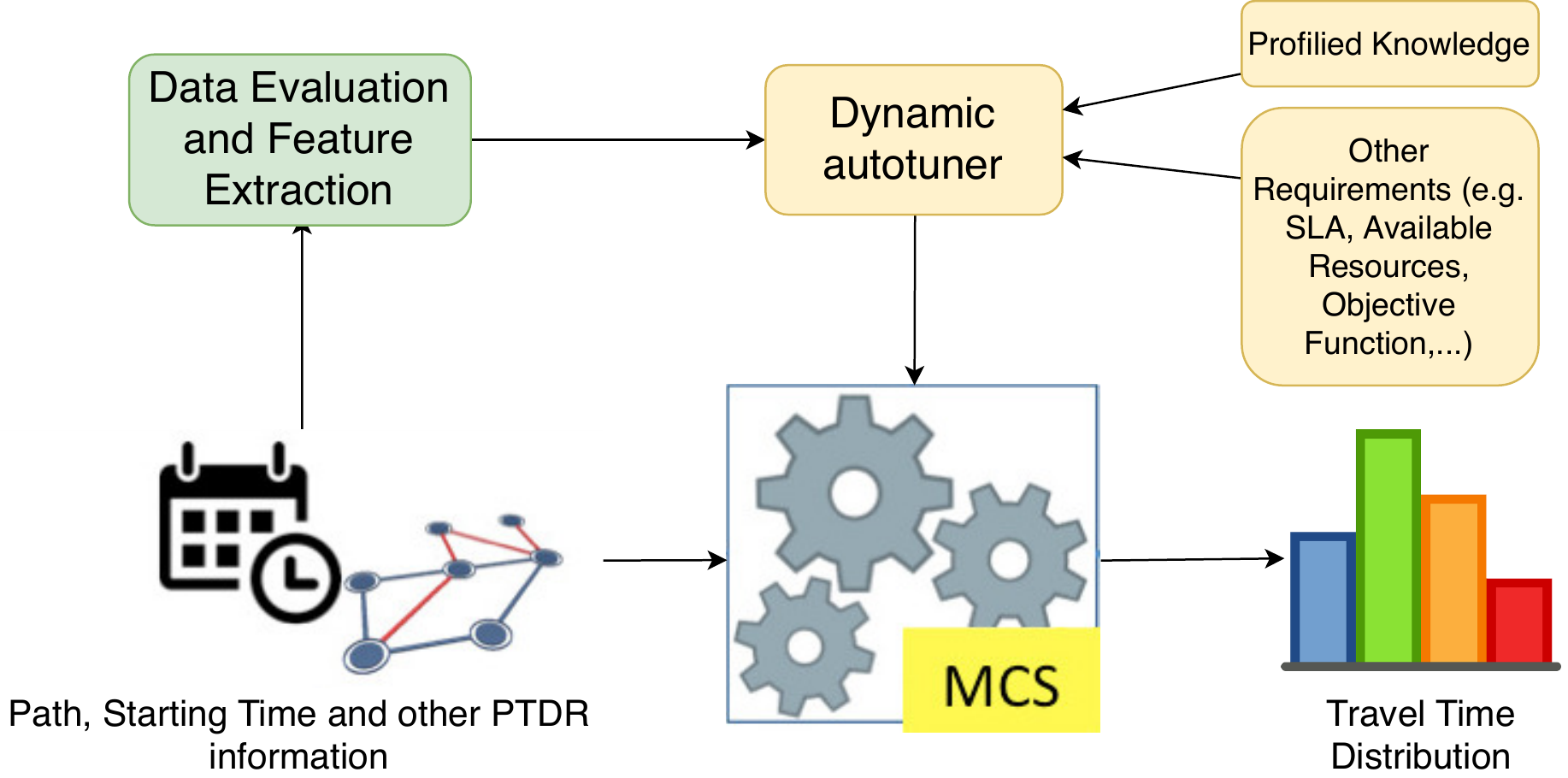}
    \caption{The proposed adaptive approach for PTDR routing based on Monte Carlo simulations and dynamic choice of the number of simulations.}
    \label{fig:final}
\end{figure}

To summarize, the proposed methodology adds an adaptivity layer on top of the Monte Carlo simulation (see \prettyref{fig:final}) to quickly determine at runtime the right number of samples for each request that satisfies the required accuracy. In particular, a feature-extraction procedure estimates the unpredictability value from the input data of the request (path, starting time and segment speed-profiles). The dynamic autotuner combines this data feature with the profiled knowledge and the extra-functional requirements to configure the Monte Carlo simulation. 

\subsection{Unpredictability Feature}
\label{sec:unpredictability}
Given that the extraction of the data feature from inputs should be done at runtime, its computation should not be a costly operation. Otherwise, the benefit of speeding up the computation phase by reducing the number of Monte Carlo samples would be reduced by the data feature extraction overhead, eventually making the whole approach meaningless.

From the experimental results, we have found that a measure of the unpredictability of the path can be extracted by a simple statistical property of the set of travel times $\theta$ extracted by a quick Monte Carlo simulation: the coefficient of variation. Intuitively the more the results are spread out, the more the route is hard to predict, thus to have a precise estimation of the distribution, and in particular the percentiles, we need a higher number of samples.

The unpredictability function is defined as $u_i = \sigma_{\theta_i}^x\slash\mu_{\theta_i}^x $ where $\sigma_{\theta_i}$ and $\mu_{\theta_i}$ are evaluated on a MCS done with the minimum number $x$ of samples allowed at runtime. It is important to note that $\sigma_{\theta_i}$ is the variance of the travel times extracted by a single Monte Carlo simulation on the minimum number of samples. % and it
We calculate the unpredictability function together with the first set of Monte Carlo samples to further reduce the overhead introduced by the data feature extraction. In particular, we will use this first short Monte Carlo run to determine if there is a need of further samples (and how many) to satisfy the error constraint.

To validate the usage of $u$ instead of $i$, we performed the Spearman correlation test \cite{Spearmanbook} between the unpredictability value and the value of $\nu_{\hat{\tau}_{i,y}^x}$ used in the calculation of the expected error for different values of $x$ and $y$ over a wide range of inputs sets $i$. In all cases, the correlation values were larger than 0.918 showing a p-value equal to 0. 
These correlations confirm our hypothesis and the p-values proves that the results are statistically significant.

\subsection{Error Prediction Function}
\label{sec:unpredictability_profiling}
To predict the expected error for a specific configuration according to the data feature $u$, we need to extract $\hat{\nu}_{\hat{\tau}_{u_i,y}^x}$ from profiling data. We run the Monte Carlo simulation several times for each configuration in terms of number of samples. In particular, we decided to use values ranging from 100 samples up to 3000. The two numbers have been derived from the observation that 100 samples is the minimum to have the estimation of the percentile for the distribution, while 3000 is the number of samples that has already been found good enough to satisfy the worst case conditions on the previous work \cite{tomis2015time}. Between the two values, we selected 2 more sampling levels corresponding to 300 and 1000, that have been derived by considering that the Monte Carlo error decreases as $1/\sqrt[]{n}$ \cite{RePEc:eee:jfinec:v:4:y:1977:i:3:p:323-338}. Thus in our case at each sampling level, we have that the error is almost halved.

We run each set of Monte Carlo simulations with the same configuration in terms of the number of samples on a large set of inputs $i$ (i.e. roads, starting time ...), and we extract $\nu_{\hat{\tau}_{u_i,y}^x}$ and $u_i$ from each single configuration.
Then we create a predictor $\hat{\nu}_{\hat{\tau}_{u_i, y}^x}$ as the \emph{quantile regression} \cite{Koenker2005QR} over the extracted data. The use of quantile regression enhances the robustness of the model in the context of its use. Indeed, we are not interested in predicting an average value as final result, but we want to use it for the inequality formula $\hat{\nu}_{\tau_{u_i,y}^x} \leq \frac{\epsilon}{n(CI)}$. In this case, a higher value of the quantile with respect to 50$^{th}$ (the purely linear regression), guarantees a higher robustness in satisfying the previous inequality.
The quantile used for the regression is an additional parameter that can be explored to trade-off robustness and performance.

\section{Integration Flow} 
\label{sec:integ}

\begin{figure}[t]
	\includegraphics[width=\columnwidth]{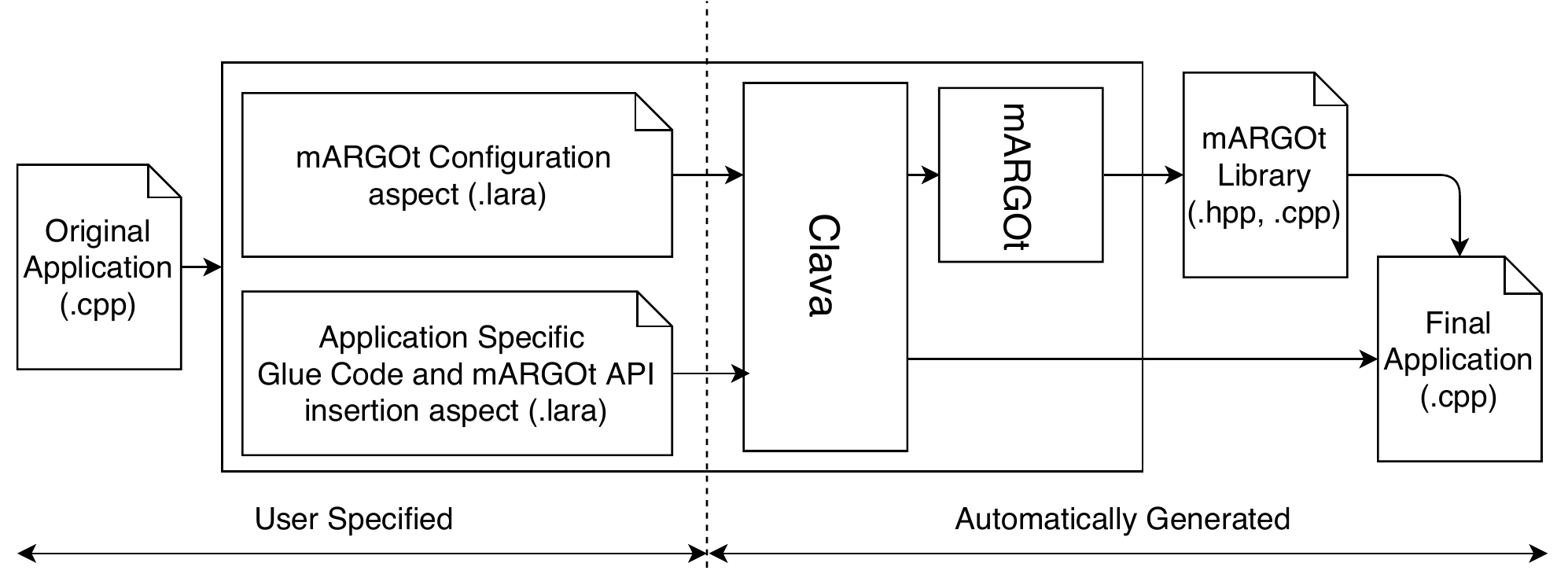}
    \caption{Integration flow outlining the two main LARA aspects and related actions: original code enrichment and autotuner configuration.}
    \label{fig:clava_mc_flow}
\end{figure}

While previous section introduces the proposed methodology from the end-user perspective, thus considering execution time and elaboration error, this section focuses on the application developer perspective by presenting the integration flow proposed to enhance the target application with limited effort. 
The proposed integration flow enforces a separation between the functional and extra-functional concerns using an Aspect-Oriented Programming Language to inject the code needed to introduce the adaptivity layer in the target source code.

On one hand we use the mARGOt \cite{gadioli2015application} framework to dynamically tune the application, thus implementing the adaptivity concepts presented in \prettyref{sec:proposedsolution} and thus transforming the target application as highlighted in \prettyref{fig:final}.
mARGOt is an open source dynamic autotuning library \footnote{Project repository: \url{https://gitlab.com/margot\_project/core}} that enhances an application with an adaptation layer.
In particular, it provides to the application the most suitable configuration of software knobs, according to application requirements and to the execution environment evolution. The configuration selection is done by relying on application knowledge gathered at profile time. 
In this context, we use mARGOt to select the number of samples that minimizes the execution time, provided that they are enough to lead to an error below a certain threshold. In particular, the selection is done by considering the unpredictability value of the current path, and using as application knowledge the design-time model described in \prettyref{sec:regressions}.

On the other hand, we hide all the complexity for code manipulation to the application developer by using LARA \cite{lara} as a language to describe user-defined strategies, and its Clava compiler \footnote{Project repository: \url{https://github.com/specs-feup/clava}} for source code analysis and transformation. 
LARA is a Domain Specific Language inspired by Aspect-Oriented Programming concepts. It allows a user to capture specific points in the code based on structural and semantic information, and then analyze and act on those points. This produces a new version of the application, leaving the original unchanged and
separating the main functional concerns from those specified in LARA.
Clava is a C/C++ source-to-source compiler based on the LARA framework. The compilation analyses and code transformations are described in scripts written in the LARA language.

\begin{lstlisting}[language=C++, caption = Original source code before integrating the adaptivity layer, captionpos=b, float, label=lst:before,breaklines=true]
// Load data
Routing::MCSimulation mc(edgesPath, profilePath);
auto run_result = mc.RunMonteCarloSimulation(samples, startTime);
ResultStats stats(run_result);
Routing::Data::WriteResultSingle(run_result, outputFile);
return 0;
\end{lstlisting}
 
 \lstset {otherkeywords = { codedef , end , aspectdef , input ,\	var ,select , apply, throw , insert , before , replace , \ new , output , true ,begin , float  , const },}

In this work, we use Clava to perform two main tasks: first, to enrich the original source code with the required autotuner glue code, and second, to configure the autotuner library according to application requirements.
\prettyref{fig:clava_mc_flow} depicts the transformation process, from the original source code, to the final application and highlights the two main LARA aspects used.
To further clarify the evolution of the application code and related aspects, Listing 1--4,
present respectively the original source code, the two LARA aspects used to enhance the target application, and the final enriched code.

\begin{lstlisting}[language=java, caption = LARA aspect for configuring the mARGOt autotuner, captionpos=b, float, label=lst:lara1, breaklines=true]
aspectdef McConfig
  /* Generated Code Structure*/
  output codegen end

  /* mARGOt configuration */
  var config = new MargotConfig();
  var travel = config.newBlock('ptdrMonteCarlo');

  /* knobs */
  ptdrMonteCarlo.addKnob('num_samples', 'samples', 'int');
  /* data features */
  ptdrMonteCarlo.addDataFeature('unpredictability', 'float', MargotValidity.GE);
  /* metrics */
  ptdrMonteCarlo.addMetric('error', 'float');
  /* goals */
  ptdrMonteCarlo.addMetricGoal('my_error_goal', MargotCFun.LE, 0.03, 'error');
  
  /* optimization problem */
  var problem = ptdrMonteCarlo.newState('problem');
  problem.setStarting(true);
  problem.setMinimizeCombination(MargotCombination.LINEAR);
  problem.minimizeKnob('num_samples', 1.0);
  problem.subjectTo('my_error_goal', 1);
    
  /* creation of the mARGOT code generator for the following code enhancement (McCodegen aspect) */
  margoCodeGen_ptdrMonteCarlo = MargotCodeGen.fromConfig(config, 'ptdrMonteCarlo');
end 
\end{lstlisting}	

In particular, the code in \prettyref{lst:lara1} shows the aspect needed to configure mARGOt, producing an autotuning library tailored according to the application requirements.
In lines 9--16, we define the \emph{num\_samples} tunable software knob, the \emph{unpredictability} feature that we want to observe, the \emph{error} metrics and the goal (i.e. the Service Level Agreement, $error < 3\%$) that in mARGOt is a condition that can be used later to define the optimization problem.
Once the knobs, metrics, and data features have been defined, we can proceed with the creation of the multi-objective constrained optimization problem that the autotuner has to manage (lines 18--23). In mARGOt optimization problems are called states (line 19). It is because mARGOt gives the possibility to define multiple optimization problems (only one can be the \emph{default} one, line 22) and also to switch among them according to dynamic conditions.
The constraints can be generated as in line 23, where the number represents the priority of the constraint. In case of more than one constraint, if the runtime is unable to satisfy both of them, it will relax the low priority one.
Lines 21--22 define the objective function. Given that in this case the objective is the minimization of the number of samples, the aspect describes it as a linear combination (line 21) of the \emph{num\_samples} knob only by using a linear coefficient equal to 1 (line 22). mARGOt and LARA integration aspects permit to build different types of combined objective functions (e.g. linear or geometric combinations).
Finally, line 26 builds the LARA internal structure \emph{margoCodeGen\_ptdrMonteCarlo} that is then used to create the mARGOt configuration file and code generator.

\begin{lstlisting}[language=java, caption = LARA aspect for inserting the application-specific glue code (unpredictability extraction) and the required mARGOt calls, captionpos=b,float , label=lst:lara3,breaklines=true]
aspectdef McCodegen
  /* Target function, mARGOt code generator from McConfig aspect, #samples for feature extraction */
  input targetName, margoCodeGen_ptdrMonteCarlo, unpredictabilitySamples end
    
  /* Target function call identification */
  select stmt.call{targetName} end
  apply 
    /* Target Code Manipulation */
    /* Add mARGOt Init*/
    margoCodeGen_ptdrMonteCarlo.init($stmt);
    /* add unpredictability code */
    $stmt.insert before UnpredictabilityCode(unpredictabilitySamples);
    /* Add mARGOt Update */
    margoCodeGen_ptdrMonteCarlo.update($stmt);
    /* Add Optimized Call Code */
    $stmt.insert replace OptimizedCall(unpredictabilitySamples);
  end
end

/* Unpredictability extraction code */
codedef UnpredictabilityCode(unpredictabilitySamples) %{
  auto travel_times_feat_new = mc.RunMonteCarloSimulation([[unpredictabilitySamples]], startTime);
  ResultStats feat_stats(travel_times_feat_new, {});
  float unpredictability = feat_stats.variationCoeff;
}% end

/* Optimized MonteCarlo call */
codedef OptimizedCall(unpredictabilitySamples) %{
  auto run_result = mc.RunMonteCarloSimulation(samples - [[unpredictabilitySamples]], startTime);
  run_result.insert(run_result.end(), travel_times_feat_new.begin(), travel_times_feat_new.end());
}% end
\end{lstlisting}

The second aspect (shown in \prettyref{lst:lara3}) aims at integrating the proposed methodology in the target application.
It takes as input (line 3) the target function call that we want to tune, the mARGOt code generator produced by the previous aspect (\prettyref{lst:lara1}), and the number of samples needed to evaluate the unpredictability feature.
In line 6, we query the code to identify the statement (\emph{stmt}) including Monte Carlo function \emph{call} as target join point to be manipulated. 
Lines 7--17 contain the actual manipulation actions done on the selected join point $stmt$ of the target code. It is composed of mainly two different types of operations. First, to integrate the mARGOt calls for initializing the library and updating the software knob (Lines 10 and 14). Second, to \emph{insert} the glue code (LARA \emph{codedef}) for calculating the unpredictability (line 12 and lines 21--25), and to \emph{replace} the original Monte Carlo call with the optimized one that does not repeat the unpredictability samples (line 16 and lines 28--31).

\lstset{ otherkeywords = {}}
\begin{lstlisting}[language=C++, caption = Target source code after the integration of the adaptivity layer, captionpos=b, float, label=lst:after,breaklines=true]
// Load data
Routing::MCSimulation mc(edgesPath, profilePath);
auto travel_times_feat_new = mc.RunMonteCarloSimulation(100, startTime);
ResultStats feat_stats(travel_times_feat_new, {});
float unpredictability = feat_stats.variationCoeff;
if(margot::travel::update(samples, unpredictability)) {
  margot::travel::manager.configuration_applied();
}
auto run_result = mc.RunMonteCarloSimulation(samples - 100, startTime);
run_result.insert(run_result.end(), travel_times_feat_new.begin(), travel_times_feat_new.end()); 
ResultStats stats(run_result);
Routing::Data::WriteResultSingle(travel_times_new, outputFile);
return 0;
\end{lstlisting}

Overall, in this specific instance of integration, we used 53 lines of LARA to generate 221 lines of C++ code. However, the advantage cannot be only considered from a numerical point of view ($>$4x in terms of line of codes). In fact, three are the main reasons to justify this approach.
First, the user does not need not to worry about the details of the mARGOt configuration files and low-level C++ API, but can instead focus on the high-level interface available in LARA that results to be more declarative on the target problem (as shown in \prettyref{lst:lara1} and \prettyref{lst:lara3}).
Second, this approach reuses information between the integration's several steps.
There is mARGOt-specific information that should be provided by the user in several places like the configuration files and when using the autotuning API (e.g. the name of the autotuner block and the knobs and data features).
By using the high-level LARA aspects, users only define this information once, saving time and possibly resulting in fewer production errors.
Third, this approach leverages on a separation of concerns between the original code (functional description) and the autotuning code (extra-functional optimization). All the extra-functional optimizations, including problem definition (optimization targets and constraints), are kept separated and users do not have to modify the original source. In this way, the original developer does not need to be involved with all the optimization process and tools, thus permitting the functional development and extra-functional optimization to run in parallel.

\section{Experimental Results}
\label{sec:experiment}

In this section, we show the results of applying the proposed methodology to the PTDR algorithm. 
The platform used for the experiments is composed by several nodes based an the Intel Xeon E5-2630 V3 CPUs (@2.8 GHz) with 128 GB of DDR4 memory (@1866 MHz) on a dual channel memory configuration.
First, we show the results of the model training for estimating the expected error (see Section \ref{sec:regressions}). Then, in Section \ref{sec:validation} we validate the approach by verifying the respect of the error constraint $\epsilon$. We compare the proposed approach with respect to the original version that takes a static decision on the number of samples (see Section \ref{sec:exp:comparison}). Finally in Section \ref{sec:exp:overhead} we discuss the overhead introduced, while in Section \ref{sec:exp:BOH} we evaluate the optimization impact when considering the entire navigation service at system-level. 

\subsection{Training the Model}
\label{sec:regressions}
\begin{figure*}[t!]
\subfloat[Quantile regression using the 50$^{th}$ perc.]{\includegraphics[width=0.67\columnwidth]{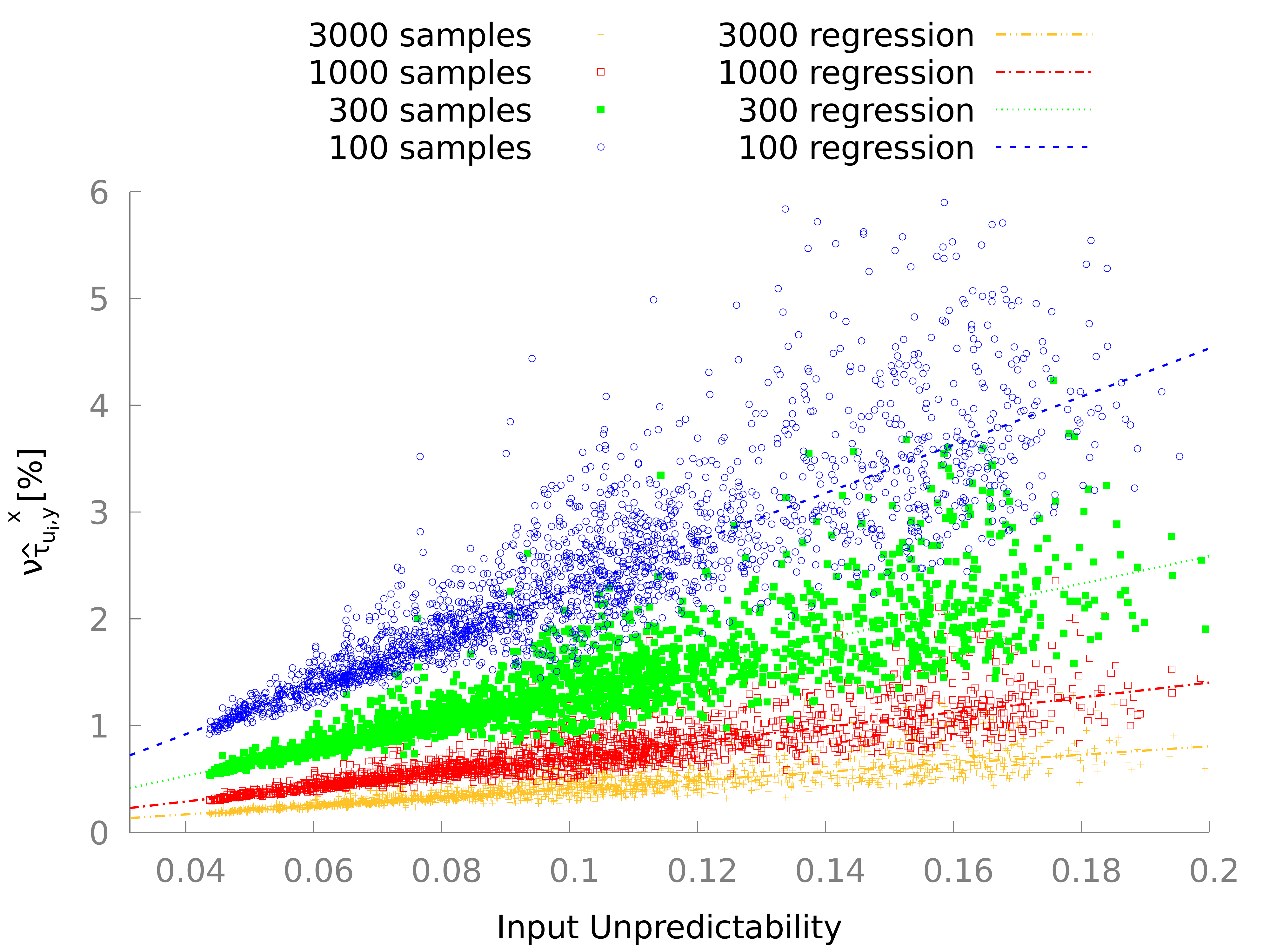}
\label{fig:regression50}}
\hfill
\subfloat[Quantile regression using the 75$^{th}$ perc.]{\includegraphics[width=0.67\columnwidth]{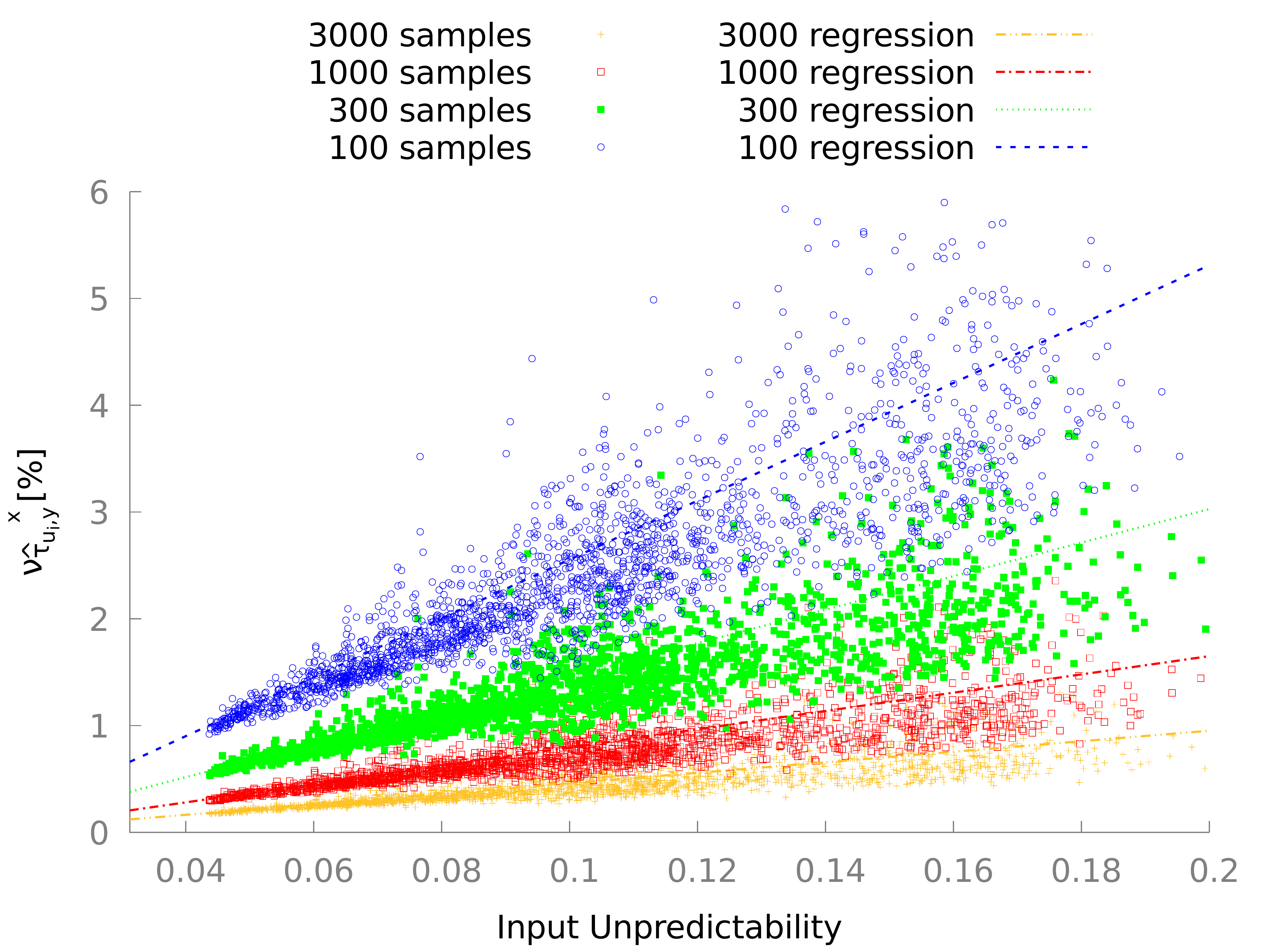}
\label{fig:regression75}}
\hfill
\subfloat[Quantile regression using the 95$^{th}$ perc.]{\includegraphics[width=0.67\columnwidth]{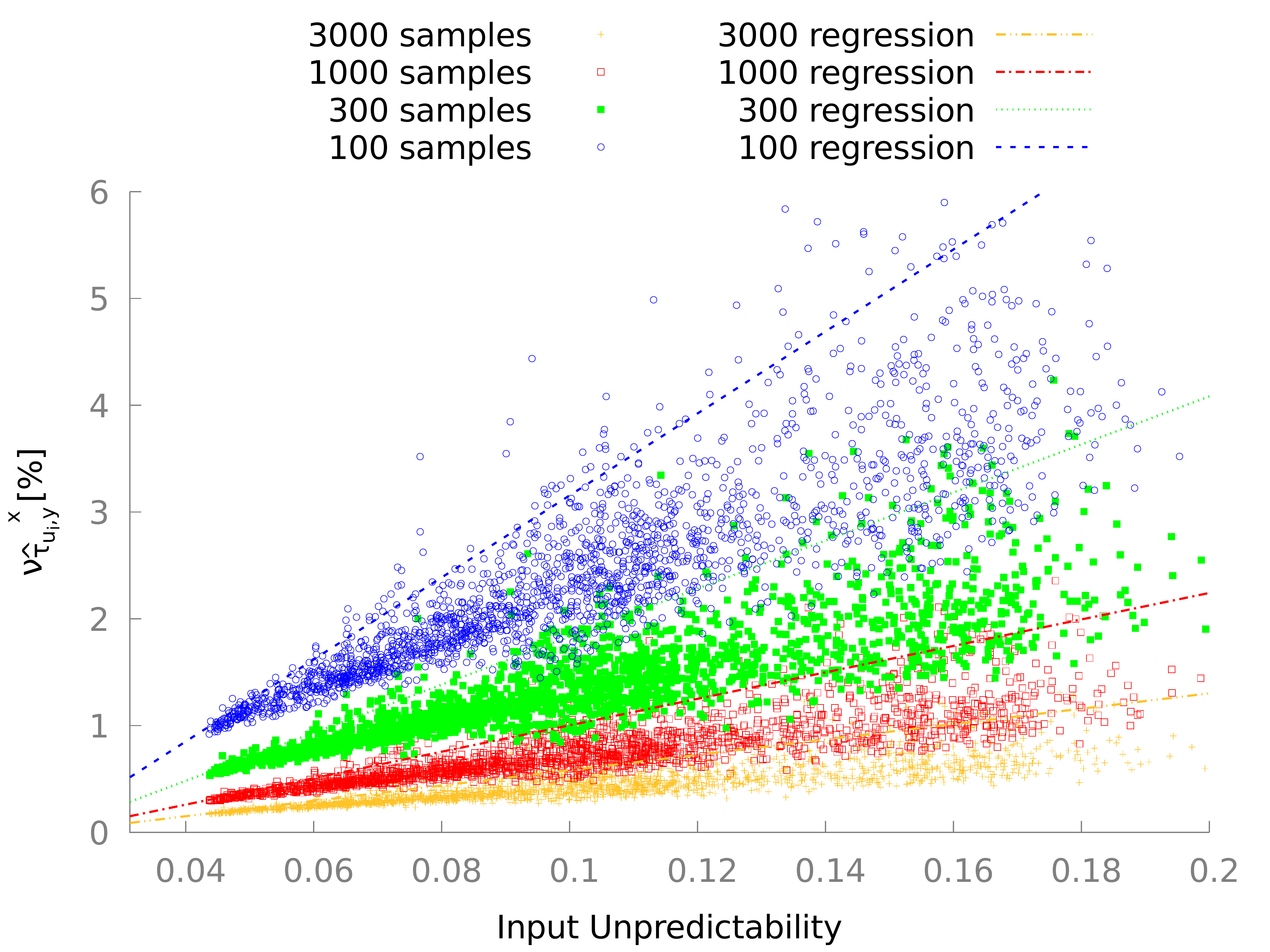}
\label{fig:regression95}}
\hfill
\caption{Training of the error model by using different number of samples and quantile regressions}
\label{fig:regressions}
\end{figure*}

The first actual phase of the methodology is done off-line and it consists of training the error model ($\hat{error}_{i,y}^x$) presented in Section \ref{sec:unpredictability_profiling} by using a different number of samples. 
For training the quantile regression, we used profiling data extracted by running the PTDR algorithm on a training set. 
This training dataset has been built using random requests done on 300 different paths across the Czech Republic in different time-slots, thus considering different speed-profiles for each segment of the paths.
All these requests have been done for all the 4 levels of sampling used in this paper (i.e. 100, 300, 1000 and 3000, as described in Section \ref{sec:unpredictability_profiling}).
The output of the model training is represented in \prettyref{fig:regressions}. The points in the three plots represent the results obtained from the profiling runs. The lines represent the quantile regression lines, thus the model that will be used at runtime. 
The three sub-figures are different in terms of the quantile value used for the regression. \prettyref{fig:regression50},  \prettyref{fig:regression75} and  \prettyref{fig:regression95} represents the regression results using respectively the 50$^{th}$, 75$^{th}$ and 95$^{th}$ quantile.
We can see that the three regressions are slightly different since we pass from a more permissive one in \prettyref{fig:regression50}, where almost half of the points are below the corresponding regression lines, to the most conservative one in \prettyref{fig:regression95} where only a few points are above. Analyzing in depth the data, we can see that the coefficients of the lines of the quantile regression are almost doubled passing from 75th to 95th percentile (e.g. for 100 samples, the coefficients pass from 0.27 to 0.38, while for 3000 samples they pass from 0.049 to 0.071).

The extracted models are now ready to be used at run-time by the dynamic autotuner to select the minimum number of samples that satisfy the error constraint for the given unpredictability value. The results shown in \prettyref{sec:validation} will demonstrate the effectiveness of the proposed method at runtime.

\subsection{Validation Results}
\label{sec:validation}

\begin{figure*}[t!]
\centering
\centering
\subfloat[Quantile regression using the 50$^{th}$ perc.]{\label{03:a}\includegraphics[width=0.65\columnwidth]{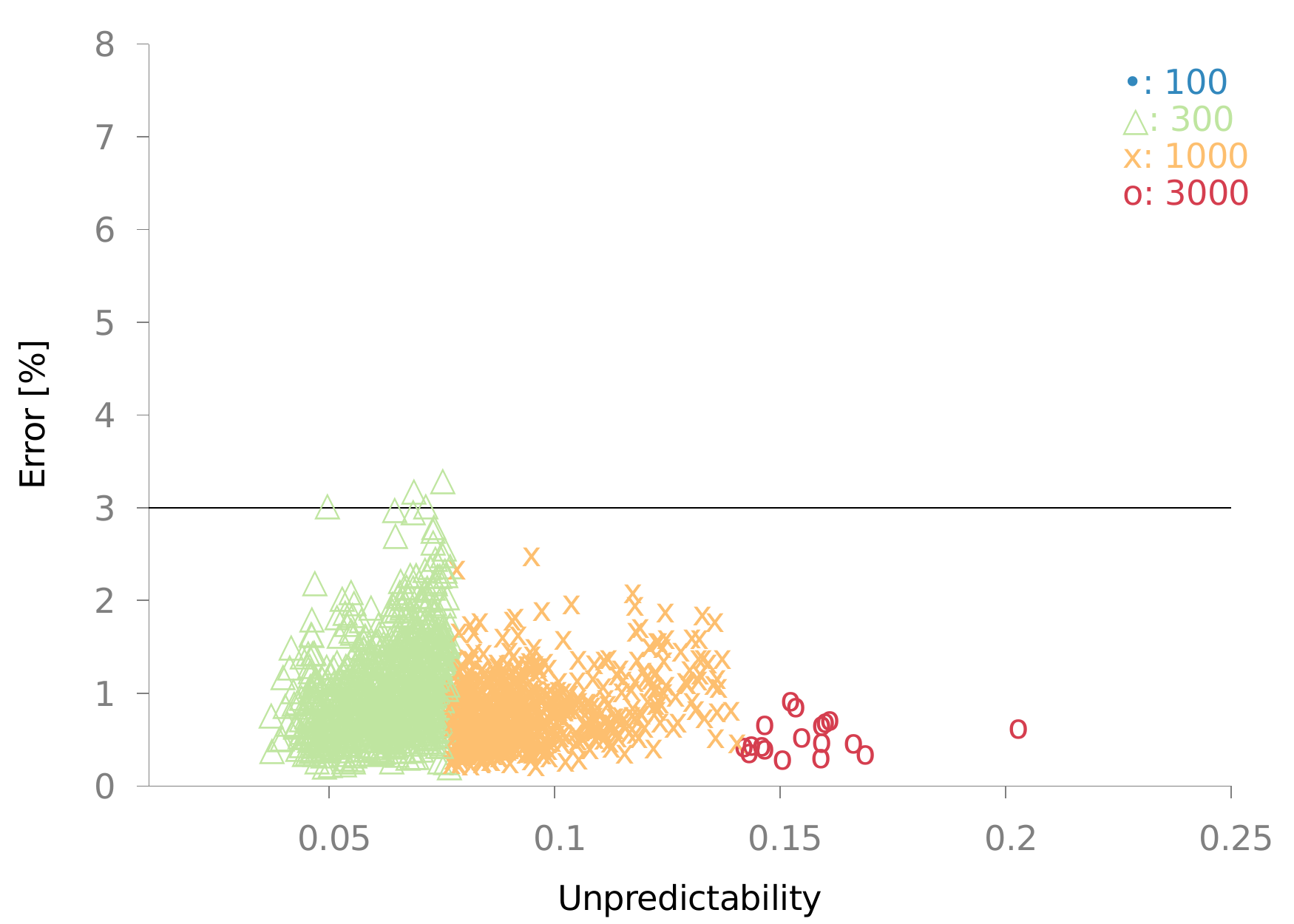}}
\hfill%
\centering
\subfloat[Quantile regression using the 75$^{th}$ perc.]{\label{fig:03b}\includegraphics[width=0.65\columnwidth]{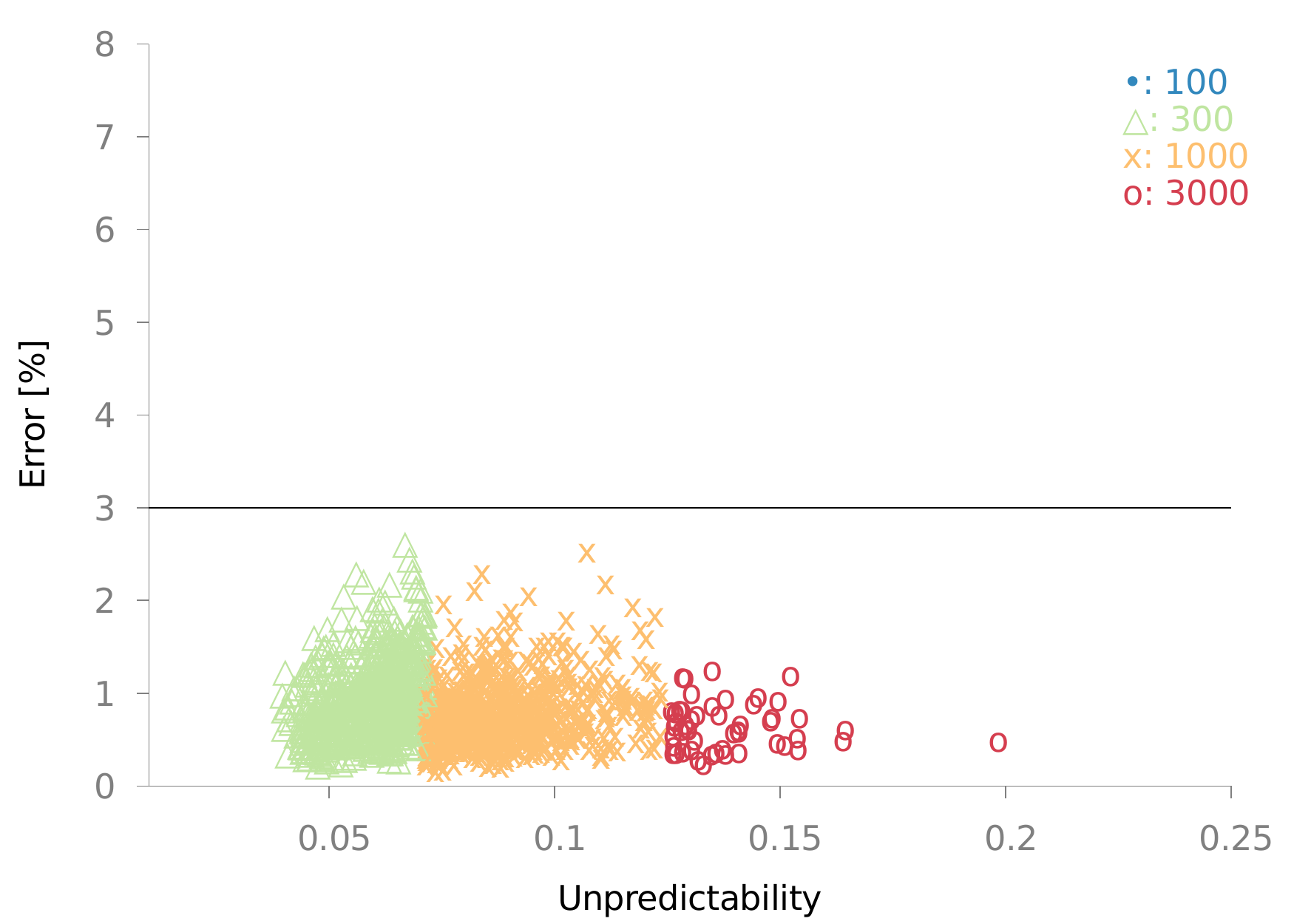}}
\hfill%
\centering
\subfloat[Quantile regression using the 95$^{th}$ perc.]{\label{03:d}\includegraphics[width=0.65\columnwidth]{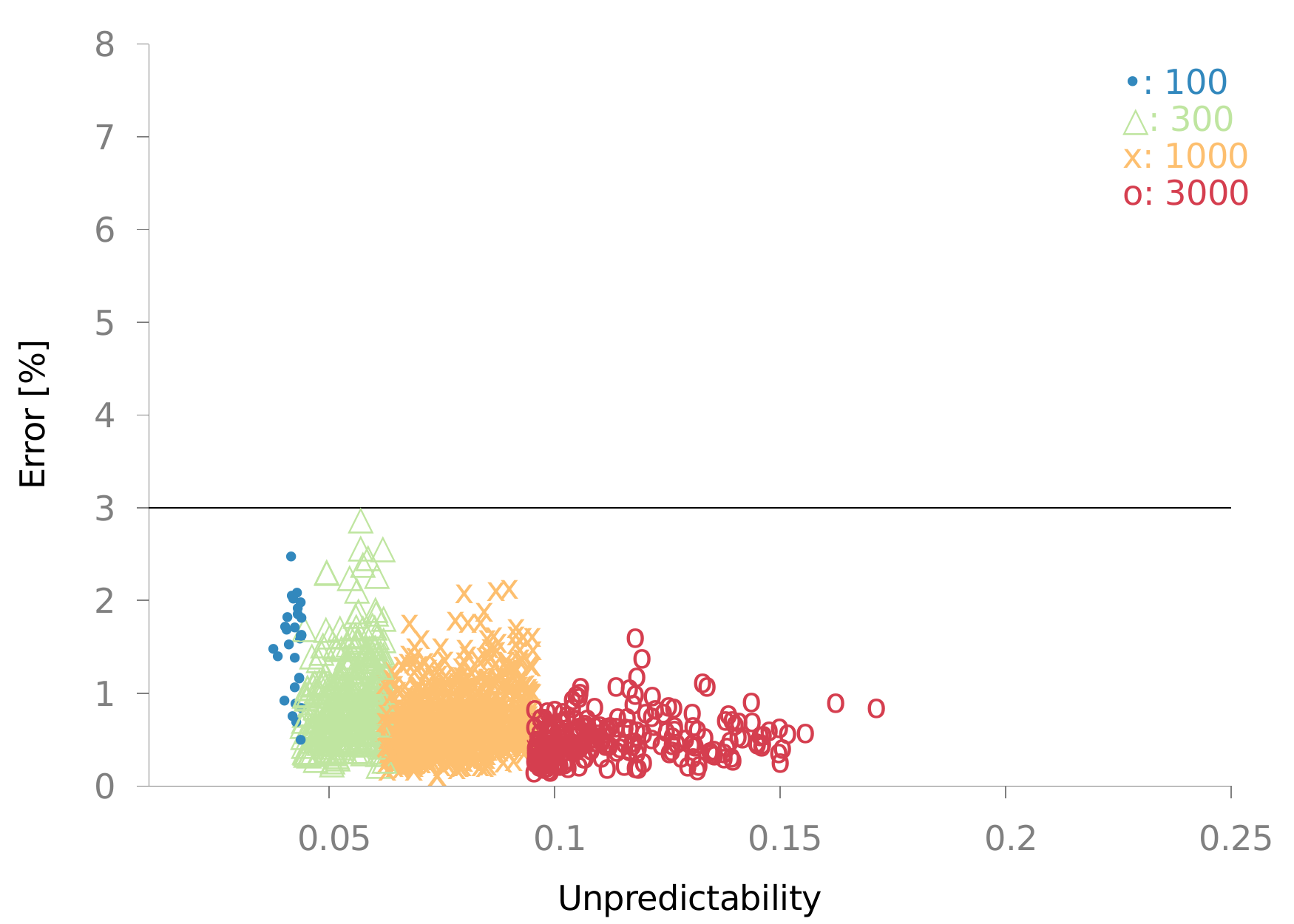}}
\caption{Validation of the proposed approach by using 3\% as target error and different percentiles for the quantile regression}
\label{fig:validation_1}
\end{figure*}

\begin{figure*}[t!]
\centering
\centering
\subfloat[Quantile regression using the 50$^{th}$ perc.]{\label{03:a}\includegraphics[width=0.65\columnwidth]{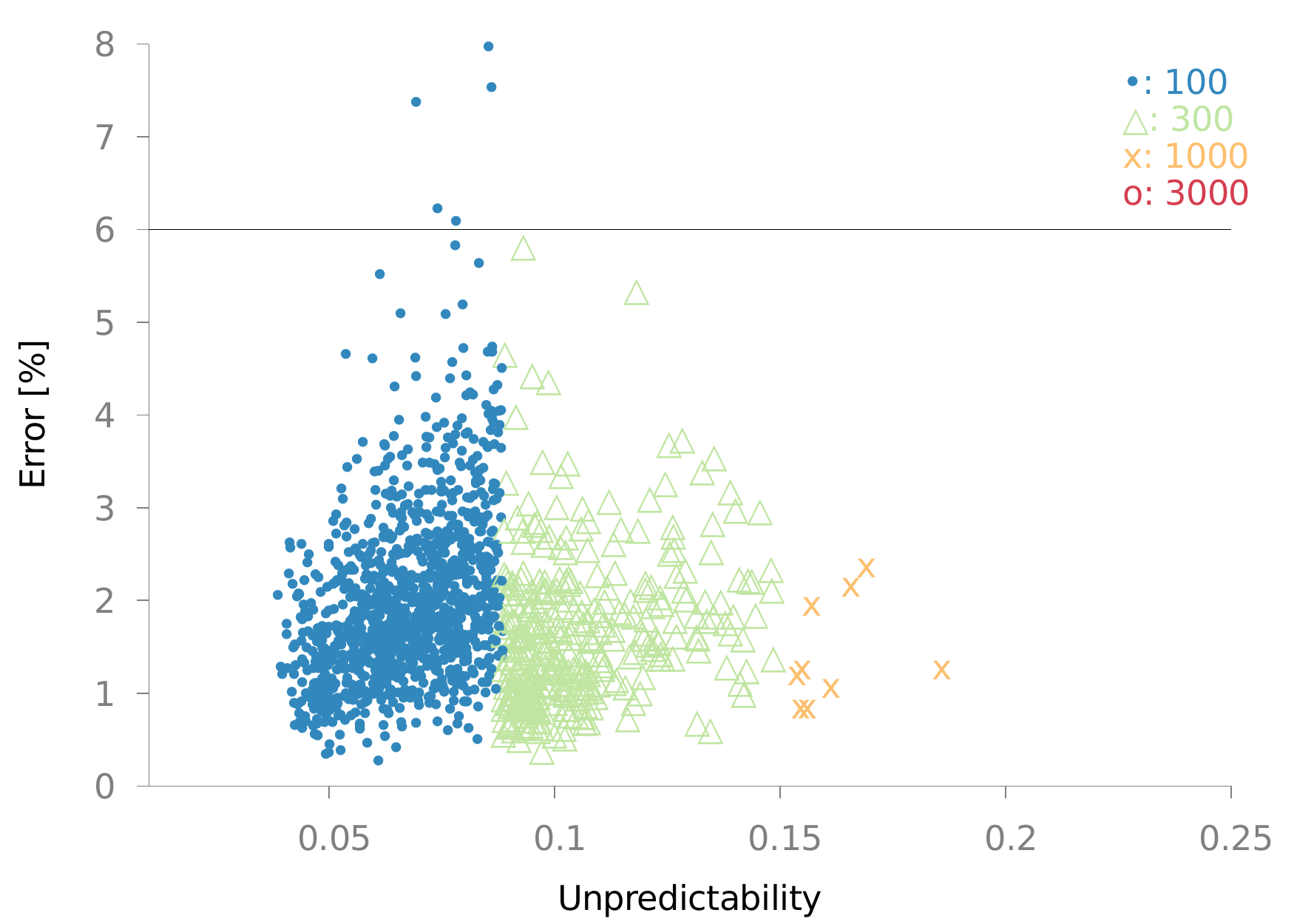}}
\hfill%
\centering
\subfloat[Quantile regression using the 75$^{th}$ perc.]{\label{fig:03b}\includegraphics[width=0.65\columnwidth]{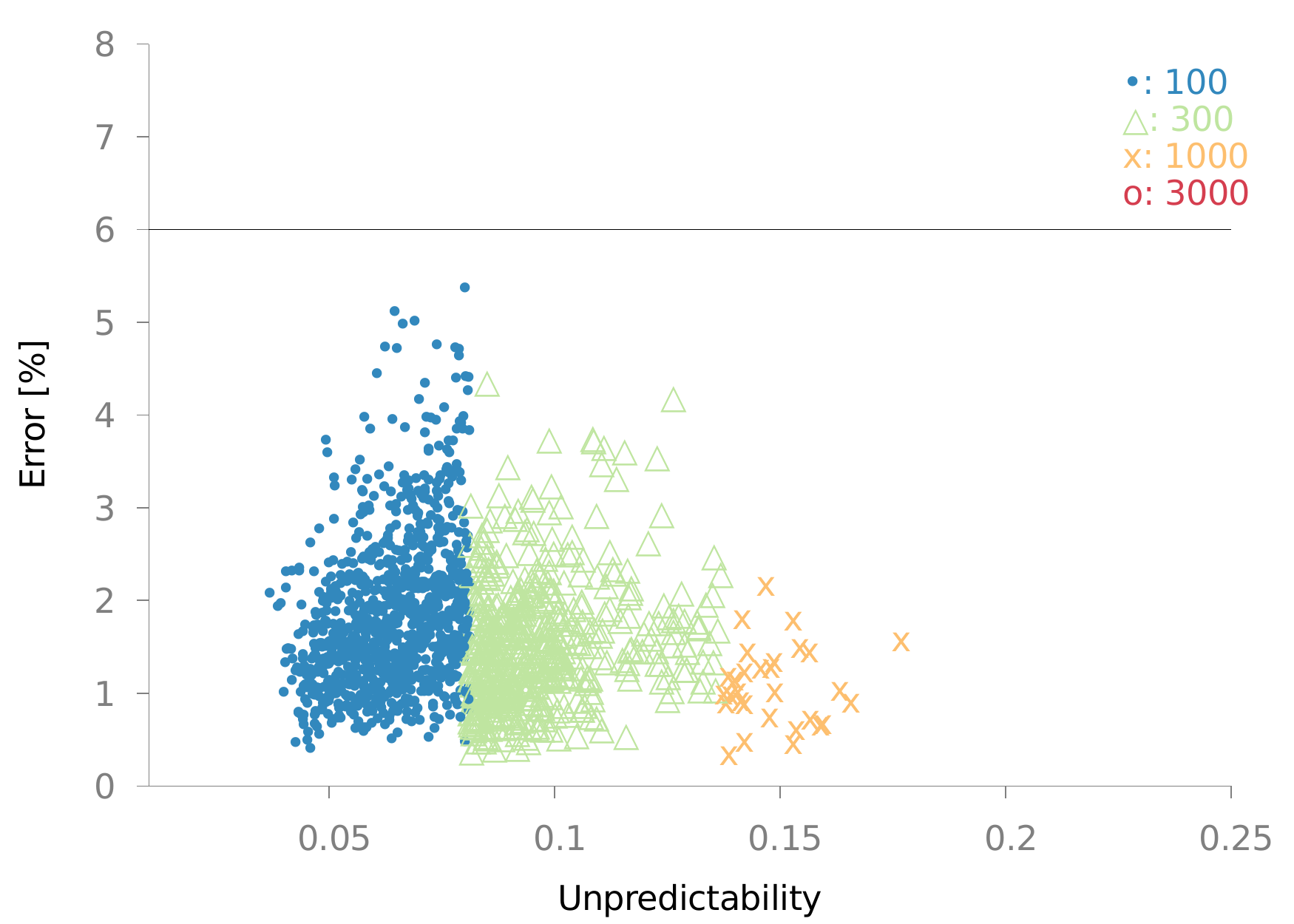}}
\hfill%
\centering
\subfloat[Quantile regression using the 95$^{th}$ perc.]{\label{03:d}\includegraphics[width=0.65\columnwidth]{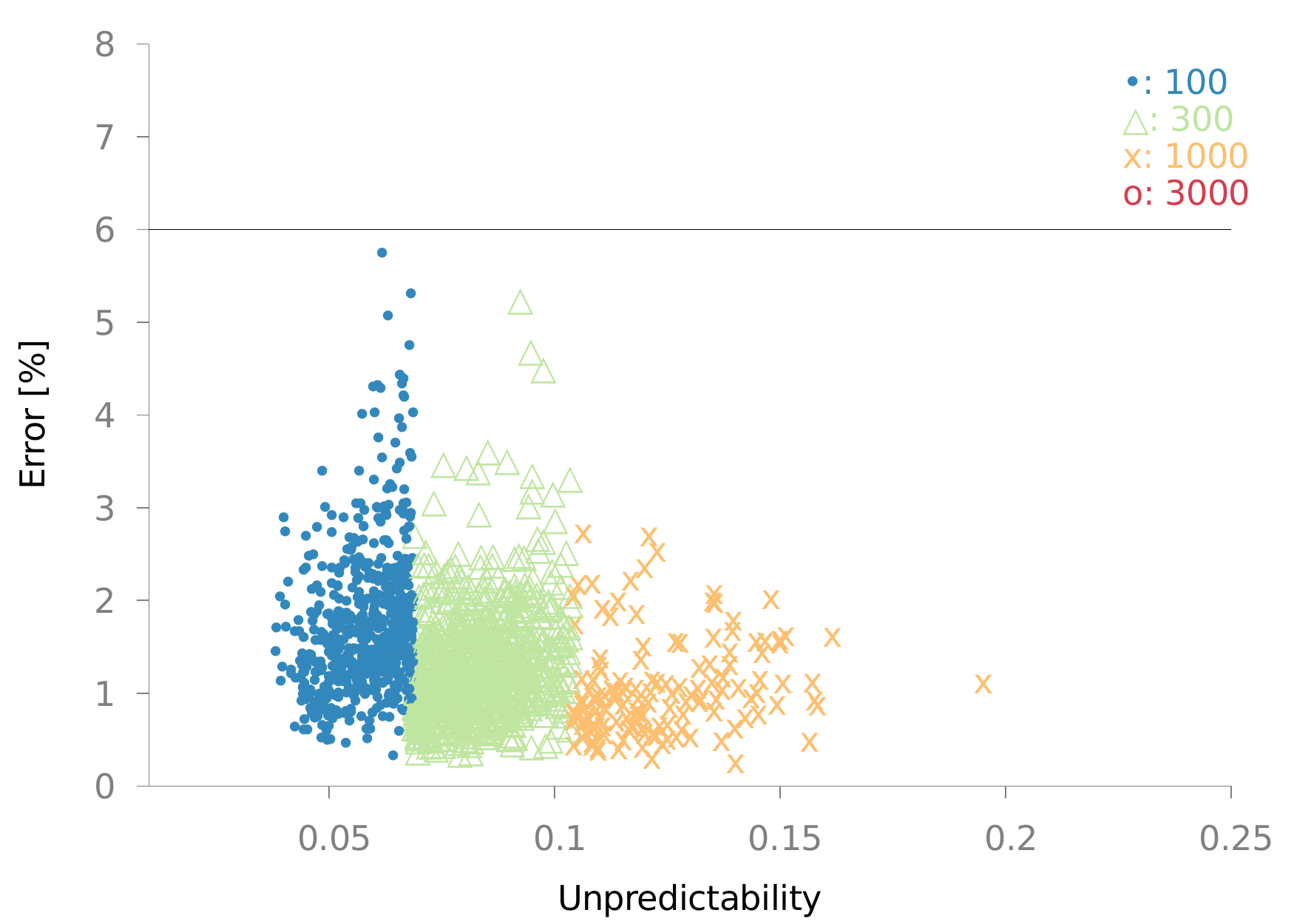}}

\caption{Validation of the proposed approach by using 6\% as target error and different percentiles for the quantile regression}
\label{fig:validation_2}
\end{figure*}

The set of validation results presented in this section are reported to demonstrate how the dynamic tuning of the number of samples satisfies the error constraints. 
For doing this, we randomly generated 1500 requests to the enhanced PTDR module for routes on the Czech Republic at different starting times. These requests are different from the ones used in the training phase of the model.
We validate the approach by using three different quantile regressions (on 50$^{th}$, 75$^{th}$ and 95$^{th}$ quantile), two different target errors $\epsilon$ (3\% and 6\%) and a confidence interval (CI) for the error constraint equal to 99\% (i.e. $n(99\%)$ = 3). 
The error has been derived by considering a run of the Monte Carlo simulation on the same input set by using 1 million of samples, thus enough to be considered a good estimation of the actual travel time distribution. Then, we selected as error the maximum between different key percentiles: 5$^{th}$, 10$^{th}$, 25$^{th}$, 50$^{th}$, 75$^{th}$, 90$^{th}$ and 95$^{th}$ percentile. 

The results are reported in \prettyref{fig:validation_1} and \prettyref{fig:validation_2} respectively for an error constraint $\epsilon$ equal to 3\% and 6\%. The two figures show the error results for each run with respect to the unpredictability feature extracted on the path. Each dot in the plots represents a PTDR request, while its shape depends on the number of samples used for the Monte Carlo simulation.
In most of the cases, the actual error is below the target error. As it was expected considering the same value of the error constraint $\epsilon$, the more conservative is the quantile regression, the less are the points that violate the constraint error. For the data we processed, in all cases the number of times the error constraint is not respected is within the selected CI (99\%). 
At the same time, moving from a less conservative quantile regression (e.g. 50$^{th}$ percentile) towards a more conservative one (e.g. 95$^{th}$ percentile), it is possible to note how the threshold values for selecting the same number of samples shifts to the left.
As an example, by considering an error constraint $\epsilon = 3\%$ (see \prettyref{fig:validation_1}), the maximum unpredictability value for having 300 samples moves from 0.075 to less than 0.06 respectively when considering the quantile regression from the 50$^{th}$ percentile, up to the 95$^{th}$ quantile. Similar is the case when we consider an error constraint $\epsilon = 6\%$ (see \prettyref{fig:validation_2}), where the same threshold moves from an unpredictability of 0.15 to 0.14 and 0.11 when using the 50$^{th}$, the 75$^{th}$ and the 95$^{th}$ as quantile value for the regression. 
Finally, it is clearly visible the difference in terms of number of samples between the two cases with different $\epsilon$ values. Indeed, while for $\epsilon$ equal to 3\% (\prettyref{fig:validation_1}) only a very small fraction of the cases use 100 samples and there is a not negligible fraction of cases where 3000 samples are employed. For $\epsilon$ equal to 6\% ( \prettyref{fig:validation_2}) in some cases only 100 samples are required.

\subsection{Comparative Results with Static Approach}
\label{sec:exp:comparison}
\begin{figure}[t]
\centering
\includegraphics[width=0.9\columnwidth]{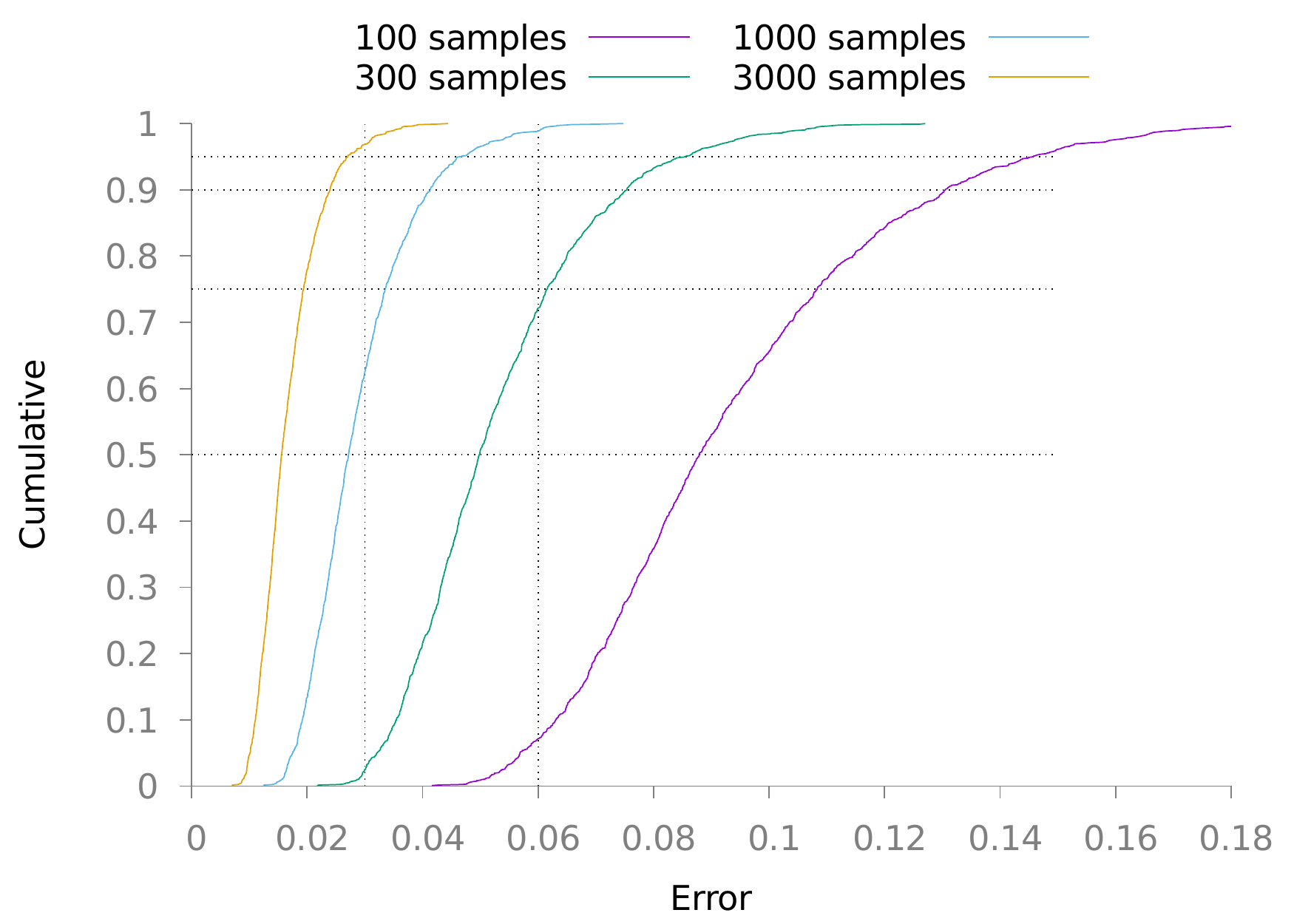}
\caption{Cumulative distribution of the error by using different numbers of samples over the training set}
\label{fig:cumulative}
\end{figure}

In this subsection, we demonstrate the advantages obtained by using the proposed approach with respect to the baseline version \cite{tomis2015probabilistic} where the number of samples are defined \emph{a priori}.
To provide a fair comparison, we extracted the number of samples to be used for the baseline version by using the training dataset.

For the 4 levels of sampling used in this paper (i.e. 100, 300, 1000 and 3000, as described in Section \ref{sec:unpredictability_profiling}), we analyzed the cumulative distributions of the expected error (see \prettyref{fig:cumulative}). 
We selected the minimum sampling level that passes a certain threshold of the cumulative value before reaching the error constraint value $\epsilon$. This threshold value has almost the same \emph{robustness} meaning of the quantile regression value used in our approach. In the following, we are going to compare the proposed approach where the quantile regression model has been built over a certain percentile, with a static tuned version where the same percentile has been used as the threshold for the cumulative.
If we use for the proposed approach the quantile regression at 95\% we compare with the statically tuned version where the number of samples has been defined looking at the cumulative curve that reaches at least 95\% before to the target error constraint.
In particular looking at \prettyref{fig:cumulative}, we can notice that for an error constraint $\epsilon=6\%$ the static tuning is set to 1000 samples for the entire percentile interval between 72$^{th}$ and 98$^{th}$, while for values larger than 98$^{th}$ and smaller than 72$^{th}$ percentile (down to 7$^{th}$) we have to consider respectively the configuration using 3000 and 300 samples.
On the other side for $\epsilon=3\%$ we select 3000 samples within the percentile interval 72$^{th}$-97$^{th}$, 1000 samples for percentile values smaller than 72$^{th}$ (down to 5$^{th}$), while we need more than 3000 samples if the request is very tight on a percentile larger than 97$^{th}$.

\begin{table}[t]
\centering

\caption{Average number of samples for the validation set using different quantile regression values (columns) and different error constraints. The results are reported for the \emph{baseline} and proposed \emph{adaptive} versions.}
\begin{tabular}{cr||c|c|c}
&&\multicolumn{3}{c}{Average Number of Samples}\\

$\epsilon$ & & $50^{th}$ perc. & $75^{th}$ perc.   & $95^{th}$ perc. \\
\hline \hline
\multirow{ 2}{*}{3\%} &baseline & 1000 & 3000  & 3000 \\
&adaptive & 632 (-36\%)& 754 (-74\%)&  1131 (-62\%)\\
\hline 

\multirow{ 2}{*}{6\%} &baseline& 300 & 1000 & 1000 \\
&adaptive & 153  (-49\%)& 186 (-81\%) & 283 (-71\%)\\

\end{tabular}
\label{tab:samples}
\end{table}

\prettyref{tab:samples} shows the comparative results obtained by using the proposed adaptive technique with respect to the original version (\emph{baseline}) with the statically defined number of samples obtained with the previously described analysis. In particular, \prettyref{tab:samples} presents the average number of samples and gain with respect to the baseline for different values of error constraint $\epsilon$ and different percentiles used to build the predictive model and for the static tuning of the baseline. The results are obtained by running a large experimental campaign over randomly selected pairs of Czech Republic routes and starting times, different from those used for the training. While the routes have been randomly selected, we used a more realistic distribution of the starting time \cite{UStrafficReport}\cite{UKtrafficReport}.

In all the considered cases, the proposed approach reduces the number of samples by at least 36\%, and up to 81\%. As expected, the average number of samples for the proposed approach is lower when we relaxed either the error constraint (i.e. 6\%) or the percentile used for building the model (e.g. 50$^{th}$ percentile). The lower gain for the configurations using the 50$^{th}$ percentile with respect to the cases using the 75$^{th}$-95$^{th}$ percentile is due to the fact that  in the former case the baseline requires a lower number of samples with respect to the latter cases (i.e. 1000 vs 3000 for $\epsilon=3\%$ and 300 vs 1000 for $\epsilon=6\%$).
Looking at the absolute numbers, it is possible to detect that even if the percentage gain seems higher with more conservative regressions (75$^{th}$-95$^{th}$), the actual average number of samples used is smaller with the more permissive quantile (50$^{th}$).

The reduction in terms of the number of samples is directly reflected on the execution time reduction since there is a linear dependency, except for the overhead introduced by the dynamic autotuner. In particular, we observed an execution time speed-up between 1.5x and 5.1x. A more detailed analysis on the overhead is presented in Section \ref{sec:exp:overhead}.

\begin{figure*}
\centering

\subfloat[Error constraint $\epsilon$=3\%]{\label{03:a}\includegraphics[width=1.8\columnwidth]{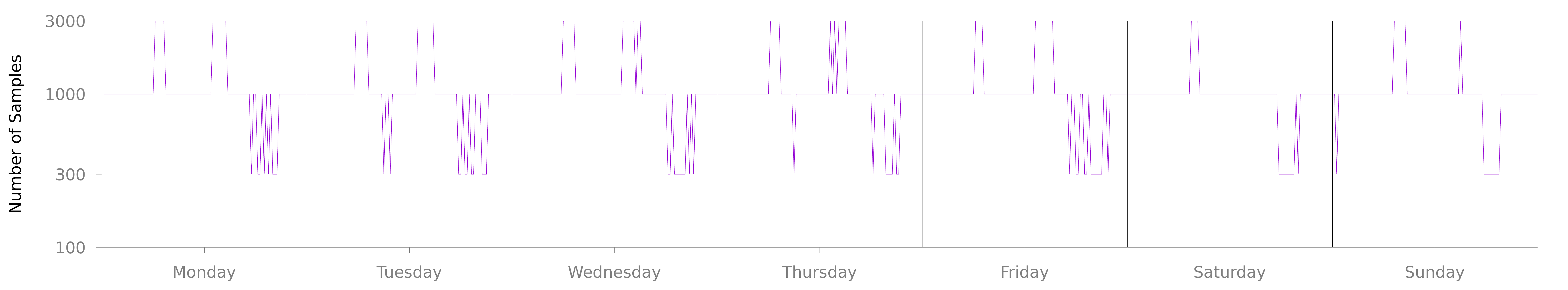}}

\subfloat[Error constraint $\epsilon$=6\%]{\label{03:d}\includegraphics[width=1.8\columnwidth]{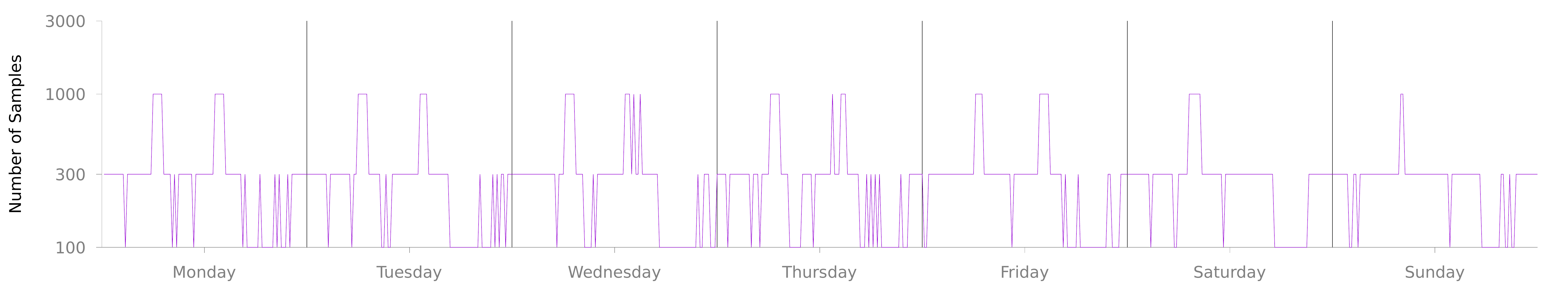}}

\caption{Number of samples selected by the proposed adaptive method when the same request is performed every 15 minutes during the entire week.}
\label{fig:validation_road}
\end{figure*}

To further show the benefits of the proposed methodology, \prettyref{fig:validation_road} shows the number of samples selected by the adaptive Monte Carlo simulation when the same request in terms of target path is performed every 15 minutes during the entire week. 
The used temporal interval is derived by the smaller time granularity ($\phi$) we had for the database containing the speed profiles.
The two plots (a) and (b) have been generated using respectively 3\% and 6\% as maximum target errors and for both experiments a quantile regression on the 75$^{th}$ percentile.

By looking at the number of samples requested by the adaptive version of the Monte Carlo simulation, we can easily recognize well-known traffic behaviors in both plots. The daily distribution on the weekdays is characterized by 2 main peaks determined by less predictable situations. The first around 7--8 am and the second around 4--5 pm. During the weekend the morning peak seems to be a bit postponed, while the afternoon one almost disappears. On the opposite, it is also visible how the evening hours result to be the most predictable ones.

This dynamic behavior that is captured by the enhanced version of the algorithm cannot be exploited by using the original (baseline) version. Following the same philosophy adopted in \prettyref{tab:samples}, the original version must be tuned by considering 3000 samples for the experiment in \prettyref{fig:validation_road}(a) ($\epsilon=3\%$) and 1000 samples for the experiment in \prettyref{fig:validation_road}(b).
In both cases, the static tuning results to be the larger number of samples selected from the proposed techniques, that instead is able to use it only when it is strictly required (e.g. during the traffic peaks).
Also considering the static tuning to the \emph{average} case (i.e. 1000 samples for the experiment in \prettyref{fig:validation_road}(a) ($\epsilon=3\%$) and 300 samples for the experiment in \prettyref{fig:validation_road}(b)) is not a viable solution. This is because there are still a lot of sampling reduction possibilities in predictable moments that will not be captured, and more important, the prediction will not be able to satisfy the algorithm output quality during the most unpredictable periods.
Finally, a fixed time-slot policy is also sub-optimal since the unpredictability strongly depends not only on the time of the request but also on the path characteristics (e.g. urban or countryside path, close or far from congested areas) and length (e.g. when it is expected the arrival in a congested area). 

\subsection{Overhead Analysis}
\label{sec:exp:overhead}

While we widely describe in \prettyref{sec:integ} how we reduced the integration overhead from the application developer point of view, this section clarifies the time-overhead introduced to obtain the proposed adaptivity.
In particular, the additional computations that we add are related to the calculation of the $\nu_{\hat{\tau}_{i,y}^x}$ and to the autotuner calls used to determine the right number of samples to be used. The initial 100 Monte Carlo samples, required to extract the data feature, are not part of the overhead given that they are reused (and thus discounted) to calculate the expected travel time (see \prettyref{lst:after}).

\begin{figure}[t]
\includegraphics[width=\columnwidth]{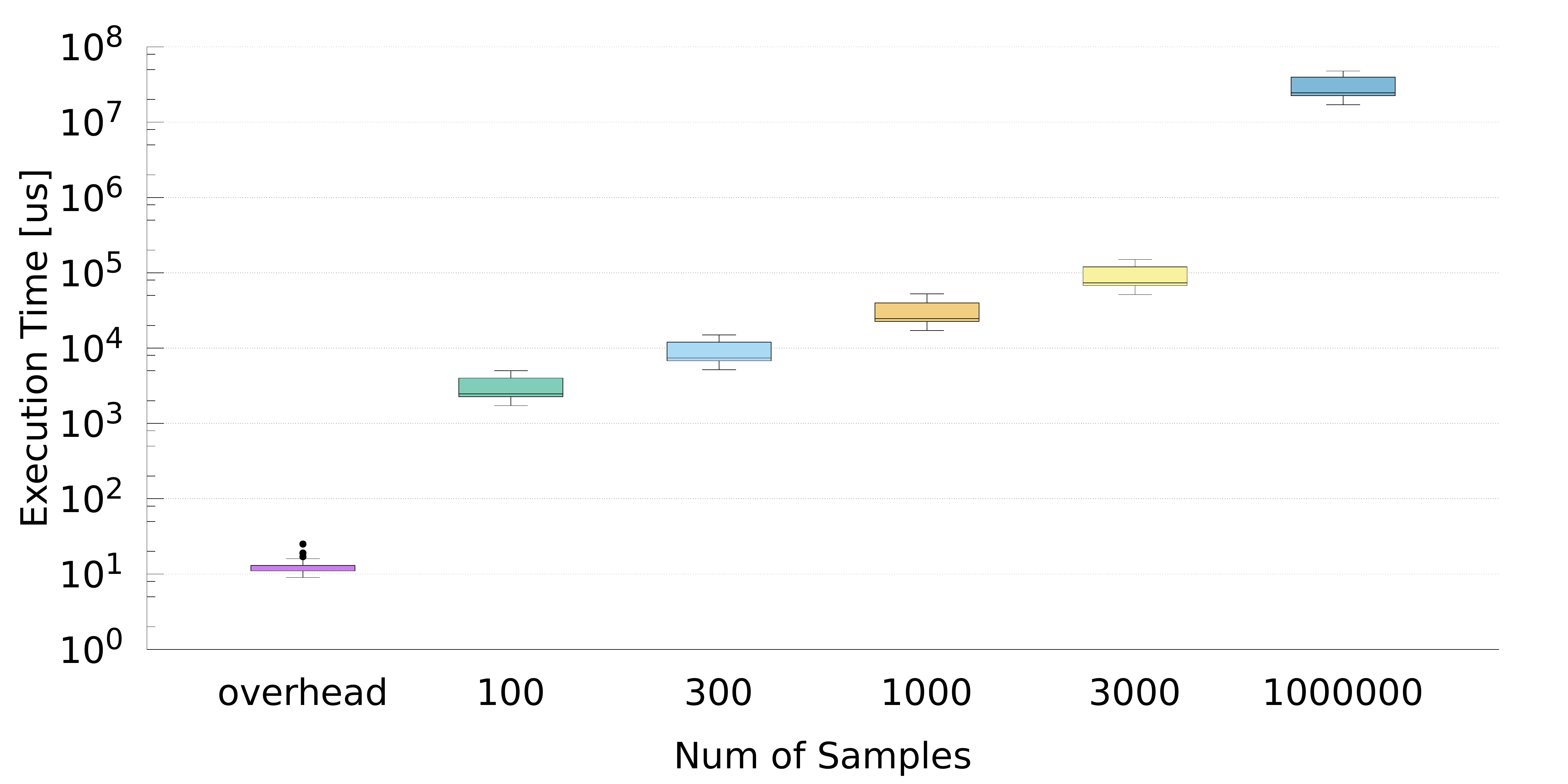}
\caption{Evaluation of the execution time overhead due to the additional code for the proposed method with respect to the target Monte Carlo simulation by varying the number of samples.}
\label{fig:overhead}
\end{figure}

\prettyref{fig:overhead} shows the overhead introduced by the proposed methodology compared to a set of Monte Carlo calculation by using a different number of samples (from 100 to 300, and 1M) over a set of paths among different locations in the three main cities in the Czech Republic.
As expected, it is evident that the execution time is strictly correlated to the number of samples used for the travel time computation.
The different paths we used are in a range between 300 and 800 segments long. When we fix the number of samples, the different number of segments is the main reason for the variability for the Monte Carlo simulation computing time.

Despite it is needed for every request, the overhead introduced by our approach is almost negligible, i.e. more than 2 orders of magnitude less than the smaller Monte Carlo simulation with 100 samples. 
In particular, we found that the execution time of the data feature extraction and mARGOt calls is comparable to the evaluation of a single sample of the Monte Carlo on a road composed of 200 segments.

\subsection{System-Level Performance Evaluation}
\label{sec:exp:BOH}

To quantify at system-level the effects of the proposed adaptive method, in this section we present an analysis done when considering that the efficient PTDR module is included in the full navigation pipeline shown in \prettyref{fig:infrastruct}. 
We built a performance model of the navigation pipeline by using the simulation environment Java Modeling Tools (JMT) \cite{Bertoli:2009:JPE:1530873.1530877}.
JMT is an integrated environment for workload characterization and performance evaluation based on queuing models \cite{Lazowska:1984:QSP:2971}. 
It can be used for 
capacity planning model simulation, workload characterization and automatic identification of
bottlenecks.
In particular, to build the simulation model of the queuing network, we considered one station for each of the modules composing the navigation pipeline and we added a fork-join unit to model the parallel PTDR evaluations of each alternative path found in the first stage. 

\begin{figure}[t]
\centering
\includegraphics[width=0.8\columnwidth]{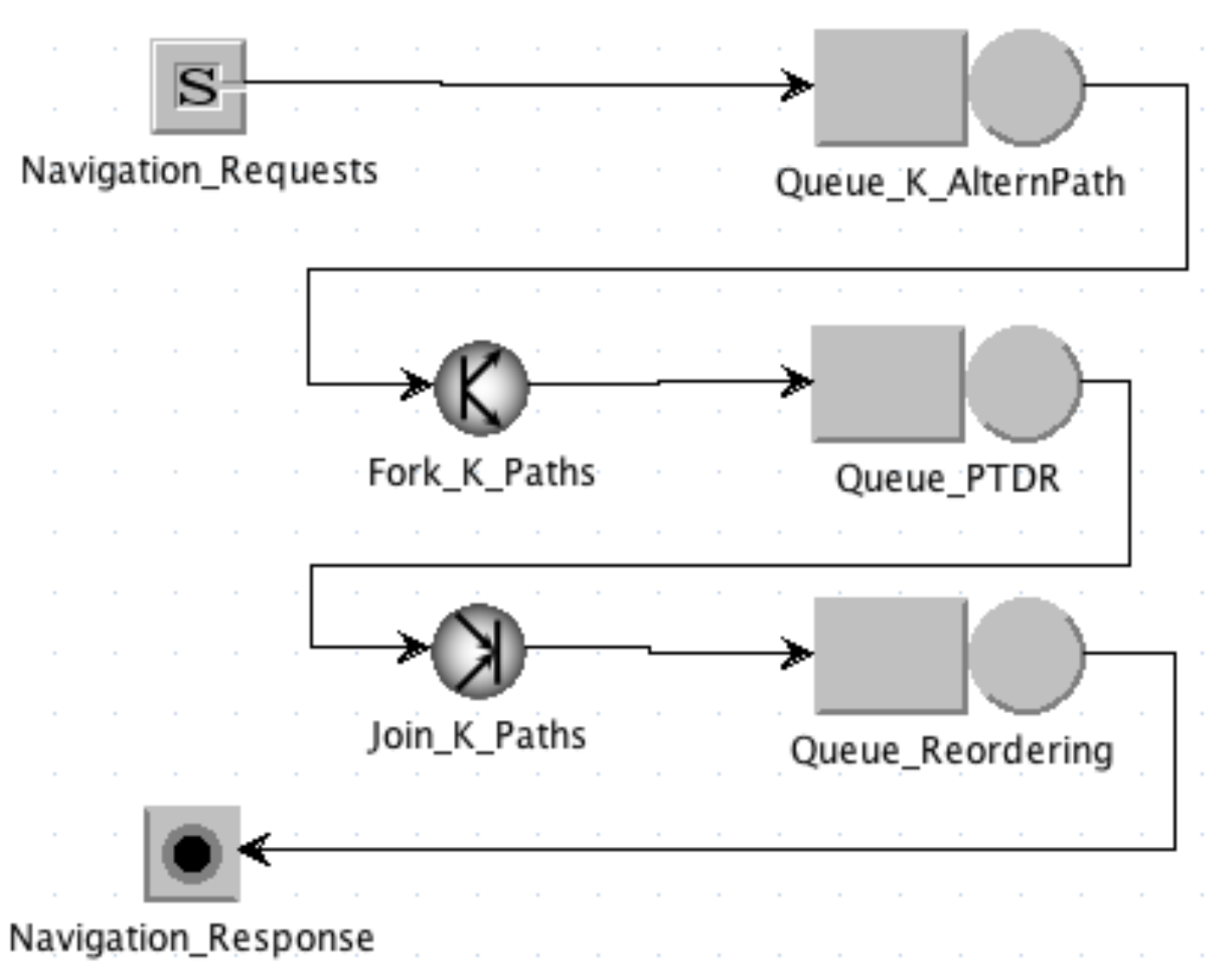}
\caption{Complete navigation pipeline modeled using JMT}
\label{fig:JMT}
\end{figure}

The model, shown in \prettyref{fig:JMT}, has been annotated with values derived by the profiling of each module (K-Alternative path, PTDR, and reordering) and considering a value for $K$ (the number of alternative paths to evaluate) equal to 10. Moreover, we made a resource allocation according to a load produced by up to 100K cars producing a request every 2 minutes. The latter number is in line with the consideration of having self-driving cars continuously connected with route planner, while the former has been derived by a simple estimation considering a Smart City such as the Milan urban area. Indeed, in this area the population is composed of around 4 Million people, every day it is estimated to have more than 5 Million trips, and only less than 50\% are done by using public transportation \cite{MilanoMobilityReport, MilanoRoadTrafficMeasures}.

Under these conditions and considering the configuration with $\epsilon=6\%$ and $95^{th} percentile$,
we found out that by adopting the proposed technique we have a 36\% reduction in terms of number of resources needed to satisfy the target workload. In particular, we can differentiate 2 cases. The first one considering the number of resources needed to satisfy the steady-state conditions, and thus that the throughput in terms of input requests should be satisfied by all the stages. 
In this case, without the proposed optimization we would have needed at least 777 computing resources (cores). Among them, 400 cores (52\% of the entire set) should be dedicated to PTDR. By applying the proposed technique only 497 cores are needed, reducing to 120 (24\% of the entire set) those required for the PTDR stage.
The second case considers a more dynamic environment where it is suggested to keep the average utilization rate of each station below 70\%. While respecting this rule of thumb \cite{gribaudoConsolidationReplication}, the distribution of the system response time (the time passing from the navigation request to the response) results to be narrow, thus being able to better react to burst of requests. 
In this second case, without the proposed optimization we would have needed 1010 cores to allocate the entire pipeline. 572 of them (57\% of the entire set) should be dedicated to the PTDR stage. By applying the proposed technique, 646 cores are enough to allocate the pipeline and out of them only 172 (26\% of the entire set) are dedicated to the PTDR.

\section{Conclusions}
\label{sec:conc}
In this paper, we presented a novel approach for dynamically select the number of samples used in a Monte Carlo simulation to solve the Probabilistic Time-Dependent Routing problem. 
The proposed method quickly samples the input data to extract an unpredictability feature used to determine in a proactive manner the number of simulation to be executed while satisfying a certain error threshold.
The runtime decision is based on a probabilistic error model -- learned offline -- correlating the unpredictability feature extracted from the data and the number of samples used by the Monte Carlo algorithm.
Experimental results demonstrated that the proposed adaptive approach for the PTDR problem is able to save a large fraction of simulations (between 36\% and 81\%) with respect to a static approach while considering different traffic situations, paths and error requirements.
Considering the entire navigation pipeline, composed also of the k-alternative path and reordering stages, the adoption of the proposed technique guarantees a significative reduction in terms of computing resources.
Finally, we adopted an aspect-oriented programming language (LARA) together with a flexible dynamic autotuning library (mARGOt) to reduce the effort necessary on the application developer for introducing the code needed to improve the execution efficiency.

\ifCLASSOPTIONcompsoc
  \section*{Acknowledgments}
\else
  \section*{Acknowledgment}
\fi

This work has been supported by European Commission under the grant 671623 FET-HPC-ANTAREX (AutoTuning and Adaptivity appRoach for Energy efficient eXascale HPC systems) and by The Czech Ministry of Education, Youth and Sports from the National Programme of Sustainability (NPU II) project ''IT4Innovations excellence in science -- LQ1602'' and by the IT4Innovations infrastructure which is supported from the Large Infrastructures for Research, Experimental Development and Innovations project ''IT4Innovations National Supercomputing Center -- LM2015070''.

\ifCLASSOPTIONcaptionsoff
  \newpage
\fi



%\bibliographystyle{IEEEtran}

\bibliographystyle{IEEEtran}
\bibliography{sample-bibliography}
%\begin{thebibliography}{1}
%H.~Kopka and P.~W. Daly, \emph{A Guide to \LaTeX}, %3rd~ed.\hskip 1em plus
%  0.5em minus 0.4em\relax Harlow, England: %Addison-Wesley, 1999.

%\end{thebibliography}


%%%%%%\begin{IEEEbiography}{Michael Shell}
%%%%%%Biography text here.
%%%%%%\end{IEEEbiography}

%%%%%%\begin{IEEEbiographynophoto}{John Doe}
%%%%%%Biography text here.
%%%%%%\end{IEEEbiographynophoto}

%\newpage

%%%%%%\begin{IEEEbiographynophoto}{Jane Doe}
%%%%%%Biography text here.
%%%%%%\end{IEEEbiographynophoto}


%\vfill


\end{document}